\documentclass[sigconf]{acmart}

\usepackage{balance}
\usepackage{todonotes}
\usepackage{multirow}
\usepackage{enumitem}

\newcommand{\query}{\mathbf{q}}
\newcommand{\doc}{\mathbf{d}}
\newcommand{\rot}[1]{}

\newcommand{\prettr}{PreTTR}

\newcommand{\edit}[1]{#1}

\newcommand{\sm}[1]{#1}
\newcommand{\fm}[1]{#1}

\AtBeginDocument{%
  \providecommand\BibTeX{{%
    \normalfont B\kern-0.5em{\scshape i\kern-0.25em b}\kern-0.8em\TeX}}}

\copyrightyear{2020} 
\acmYear{2020} 
\setcopyright{acmcopyright}\acmConference[SIGIR '20]{Proceedings of the 43rd International ACM SIGIR Conference on Research and Development in Information Retrieval}{July 25--30, 2020}{Virtual Event, China}
\acmBooktitle{Proceedings of the 43rd International ACM SIGIR Conference on Research and Development in Information Retrieval (SIGIR '20), July 25--30, 2020, Virtual Event, China}
\acmPrice{15.00}
\acmDOI{10.1145/3397271.3401093}
\acmISBN{978-1-4503-8016-4/20/07}

\hypersetup{draft}
\begin{document}

\title[Efficient Document Re-Ranking for Transformers by Precomputing Term Representations]{Efficient Document Re-Ranking for Transformers by Precomputing Term Representations}

\author{Sean MacAvaney}
\affiliation{\institution{IR Lab, Georgetown University, USA}}
\email{sean@ir.cs.georgetown.edu}

\author{Franco Maria Nardini}
\affiliation{\institution{ISTI-CNR, Pisa, Italy}}
\email{francomaria.nardini@isti.cnr.it}

\author{Raffaele Perego}
\affiliation{\institution{ISTI-CNR, Pisa, Italy}}
\email{raffaele.perego@isti.cnr.it}

\author{Nicola Tonellotto}
\affiliation{\institution{University of Pisa, Italy}}
\email{nicola.tonellotto@unipi.it}

\author{Nazli Goharian}
\affiliation{\institution{IR Lab, Georgetown University, USA}}
\email{nazli@ir.cs.georgetown.edu}

\author{Ophir Frieder}
\affiliation{\institution{IR Lab, Georgetown University, USA}}
\email{ophir@ir.cs.georgetown.edu}

\fancyhead{}


\begin{abstract}
Deep pretrained transformer networks are effective at various ranking tasks, such as question answering and ad-hoc document ranking. However, their computational expenses deem them cost-prohibitive in practice. Our proposed approach, called \prettr{} (Precomputing Transformer Term Representations), considerably reduces the query-time latency of deep transformer networks (up to a $42\times$ speedup on web document ranking) making these networks more practical to use in a real-time ranking scenario. Specifically, we precompute part of the document term representations at indexing time \edit{(without a query)}, and merge them with the query representation at query time to compute the final ranking score. Due to the large size of the token representations, we also propose an effective approach to reduce the storage requirement by training a compression layer to match attention scores. Our compression technique reduces the storage required up to \sm{95\%} and it can be applied without a substantial degradation in ranking performance.
\end{abstract}




\maketitle

\section{Introduction}

\begin{figure}
\centering
\vspace{1.5em}
\includegraphics[scale=0.65]{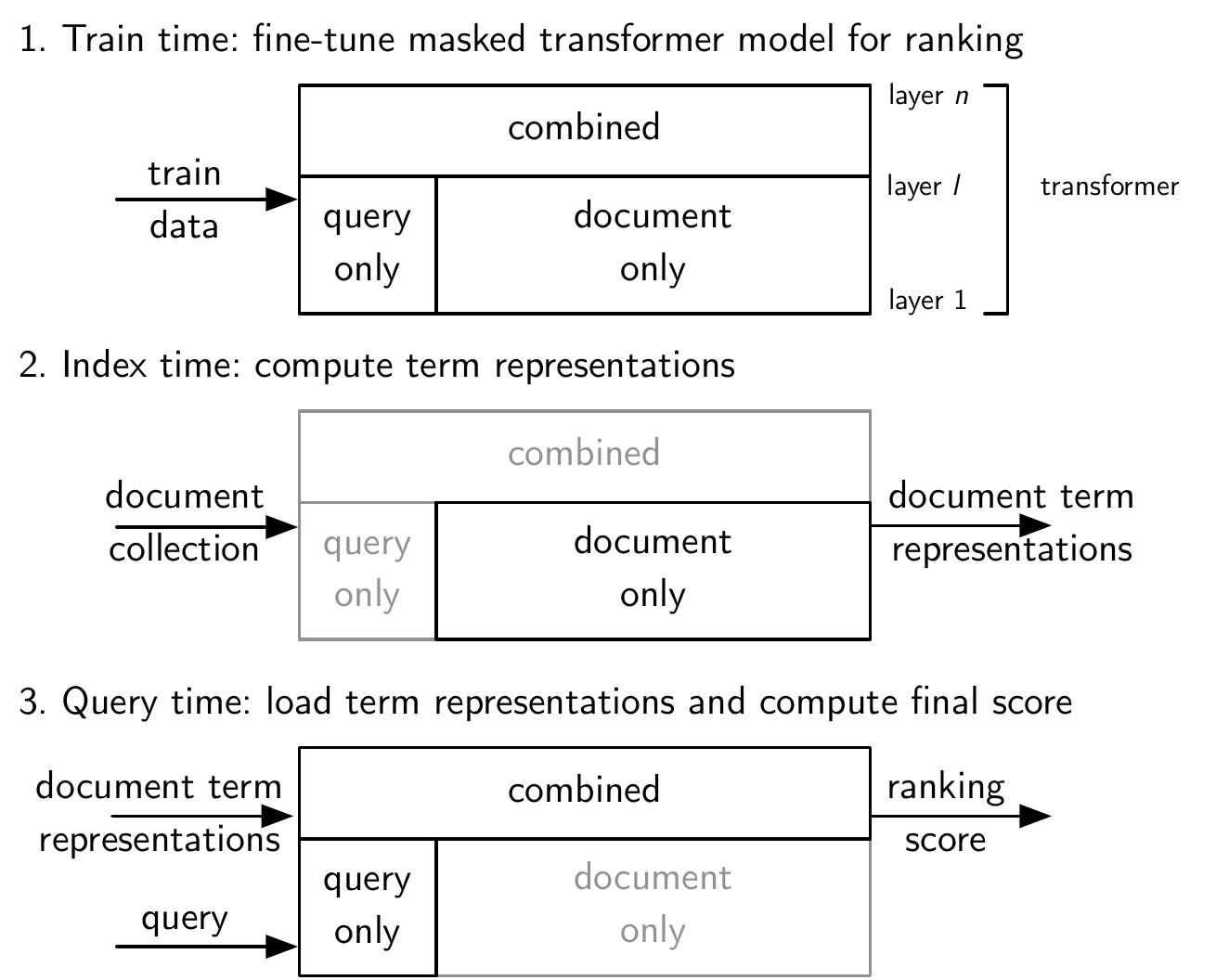}
\caption{High-level overview of \prettr. \sm{At query time, document representations (which were computed at index time) are loaded, which reduces the computational burden.}}
\label{fig:overview}
\end{figure}

Pretrained deep transformer networks, e.g., BERT~\cite{devlin-19}, have recently been transformative for many tasks, exceeding the effectiveness of prior art in many natural language processing and information retrieval tasks~\cite{Nogueira2019PassageRW,Nogueira2019DocumentEB,Yang2019SimpleAO,MacAvaney2019CEDRCE,Dai2019DeeperTU,Yang2019EndtoEndOQ}. However, these models are huge in size, thus expensive to run. For instance, in about one year, the largest pretrained transformer model grew from about $110$ million parameters (GPT~\cite{radford2018improving}) to over $8.3$ billion (Megatron-LM~\cite{Shoeybi2019MegatronLMTM}), which, when applied to IR tasks like ad-hoc retrieval, have substantial impact on the query processing performance, to the point of being impractical~\cite{MacAvaney2019CEDRCE}. We move these neural ranking models towards practicality.

Runtime efficiency is a central tenant in information retrieval, though as neural approaches have gained prominence, their running time has been largely ignored in favor of gains in ranking performance~\cite{Hofsttter2019LetsMR}. Recently, the natural language processing community has begun to consider and measure running time~\cite{Schwartz2019GreenA}, albeit mostly for reasons of environmental friendliness and inclusiveness. Chiefly, model distillation  approaches~\cite{Tang2019DistillingTK,Jiao2019TinyBERTDB,Sanh2019DistilBERTAD} are prominent, which involve training a smaller model off of the predictions of a larger model. This smaller model can then be further fine-tuned for a specific task. While this approach can exceed the performance of a smaller model when only trained on the specific task data, it inherently limits the performance of the smaller model to that of the larger model. Nevertheless, distillation is a method complementary to ours; our approach can work with a distilled transformer network. Others have explored quantization approaches to reduce model sizes, by limiting the number of bits used to represent network's parameters to 16, 8, or fewer bits. Quantization was mainly explored to make the neural networks suitable for embedded systems~\cite{Han2015DeepCC,Seo2019EfficientWQ}. We employ a basic quantization technique to reduce the storage requirements of the term representations.

We propose a method for improving the efficiency of transformer-based neural ranking models. We exploit a primary characteristic of ad-hoc ranking: an initial indexing phase can be employed to pre-process documents in the collection to improve query-time performance. Specifically, we observe that much of the term interaction at query time happens locally within either the query or document, and only the last few layers of a deep transformer network are required to produce effective ranking scores once these representations are built. Thus, documents can be processed at index time through part of the network \edit{without knowledge of the query}. The output of this partial network computation is a sequence of contextualized term representations. These representations can then be stored and used at query time to finish the processing in conjunction with the query. \edit{This approach can be trained end-to-end by masking the attention across the query and document during training time (i.e., disallowing the document from attending to the query and vice versa.)} We call this approach \prettr{} (Precomputing Transformer Term Representations). A high-level overview of \prettr{} is shown in Figure~\ref{fig:overview}.

At train time, a transformer network is fine-tuned for ad-hoc document ranking. This transformer network masks attention scores in the first $l$ layers, disallowing interactions between the query and \fm{the} document. At index time, each document in the collection is processed through the first $l$ layers, and the resulting term representations are stored. At query time, the query is processed through the first $l$ layers, and then combined with the document term representations to finish the ranking score calculation.

Since term representations of each layer can be large (e.g., $768$ float values per document term in the base version of BERT), we also propose a compression approach. This approach involves training an encoding layer between two transformer layers that produces representations that can replicate the attention patterns exhibited by the original model. We experimentally show that all these processes result in a much faster network at query time, while having only a minimal impact on the ranking performance and a reasonable change in index size. \edit{The settings of \prettr{} (amount of pre-computation, degree of compression) can be adjusted depending on the needs of the application.} These are all critical findings that are required to allow transformer networks to be used in practical search environments. Specifically, the lower computation overhead reduces query-time latency of using transformer networks for ranking, all while still yielding the substantial improvements to ranking accuracy that transformer-based rankers offer.

In summary, the contributions of the paper are the following:
\vspace{-0.3em}
\begin{itemize}[leftmargin=2em]
\item A new method for improving the efficiency of \edit{transformer-based} neural ranking models (\prettr{}). The approach exploits the inverted index to store a precomputed term representation of documents used to improve query-time performance;
\item A novel technique for compressing the precomputed term representations to reduce the storage burden introduced by \prettr{}. This is accomplished by training a compression function between transformer layers to minimize the difference between the attention scores with and without compression;
\item A comprehensive experimental evaluation of \edit{\prettr{} on multiple pre-trained transformer networks} on two public datasets, namely, TREC WebTrack 2012 and TREC Robust 2004. Our \prettr{} accelerates the document re-ranking stage by up to $42\times$ on TREC WebTrack 2012, while maintaining comparable P@20 performance. Moreover, our results show that our compression technique can reduce the storage required by \prettr{} by up to 97.5\% without a substantial degradation in the ranking performance;
\item \sm{For reproducibility, our code is integrated into OpenNIR~\cite{macavaney:wsdm2020-onir}, with instructions and trained models available at: \\ \url{https://github.com/Georgetown-IR-Lab/prettr-neural-ir}.}
\end{itemize}

\section{Related Work}
\label{sec:related}
We present an overview of neural ranking techniques, pretrained transformers for ranking, and efforts to optimize the efficiency of such networks.

\subsection{Neural Ranking}
As neural approaches have gained prominence in other disciplines, many have investigated how deep neural networks can be applied to document ranking~\cite{Huang2013LearningDS,guo-16,xiong-17,Hui2017PACRRAP}. These approaches typically act as a final-stage ranking function, via a \textit{telescoping} \sm{(also referred to as \textit{cascading}, or \textit{multi-stage}) technique~\cite{Matveeva2006HighAR,wang2011cascade}}; that is, initial ranking is conducted with less expensive approaches (e.g., BM25), with the final ranking score calculated by the more expensive machine-learned functions. This technique is employed in commercial web search engines~\cite{Rosset2018OptimizingQE}. Neural ranking approaches can broadly be categorized into two categories: \textit{representation-focused} and \textit{interaction-focused} models. Representation-focused models, such as DSSM~\cite{Huang2013LearningDS}, aim to build a dense ``semantic'' representation of the query and the document, which can be compared to predict relevance. This is akin to traditional vector space models, with the catch that the vectors are learned functions from training data. Interaction models, on the other hand, learn patterns indicative of relevance. For instance, PACRR~\cite{Hui2017PACRRAP} learns soft n-gram matches in the text, and KNRM~\cite{xiong-17} learns matching kernels based on word similarity scores between the query and the document.

\subsection{Pretrained Transformers for Ranking}
\label{ssec:pretrain}
Since the rise of pretrained transformer networks (e.g., BERT~\cite{devlin-19}), several have demonstrated their effectiveness on ranking tasks. \citet{Nogueira2019PassageRW} demonstrated that BERT was effective at passage re-ranking (namely on the MS-MARCO and TREC CAR datasets) by fine-tuning the model to classify the query and passage pair as relevant or non-relevant. \citet{Yang2019EndtoEndOQ} used BERT in an end-to-end question-answering pipeline. In this setting, they predict the spans of text that answer the question (same setting as demonstrated on SQuAD in~\cite{devlin-19}). \citet{MacAvaney2019CEDRCE} extended that BERT is effective at \textit{document} ranking, both in the ``vanilla'' setting (learning a ranking score from the model directly) and when using the term representations from BERT with existing neural ranking architectures (CEDR). \citet{Dai2019DeeperTU} found that the additional context given by natural language queries (e.g., topic descriptions) can improve document ranking performance, when compared with keyword-based queries. \citet{Yang2019SimpleAO} showed that BERT scores aggregated by sentence can be effective for ranking. Doc2Query~\cite{Nogueira2019DocumentEB} employs a transformer network at index time to add terms to documents for passage retrieval.
The authors also demonstrate that a BERT-based re-ranker can be employed atop this index to further improve ranking performance.

\subsection{Neural Network Efficiency}
\label{ssec:efficiency}
Pretrained transformer networks are usually characterized by a very large numbers of parameters and very long inference times, making them unusable in production-ready IR systems such as web search engines. Several approaches were proposed to reduce the model size and the inference computation time in transformer networks~\cite{DBLP:journals/corr/HanMD15}. Most of them focus on the compression of the neural network to reduce their complexity and, consequently, to reduce their inference time.

Neural network \emph{pruning} consists of removing weights and activation functions in a neural  network to reduce the memory needed to store the network parameters. The objective of pruning is to convert the weight matrix of a dense neural network to a sparse structure, which can be stored and processed more efficiently.  Pruning techniques work both at learning time and as a post-learning step. In the first category, Pan et al. propose regularization techniques focused at removing redundant neurons at training time~\cite{DBLP:journals/corr/PanDG16}. Alternatively, in the second category, Han et al. propose to remove the smallest weights in terms of magnitude and their associated edges to shrink the size of the network~\cite{han2015learning}. 
Conversely, our proposed approach does not change the dense structure of a neural network to a sparser representation, but it aims to precompute the term representation of some layers, thus completely removing the document-only portion of a transformer neural network (see Figure~\ref{fig:overview}).

Another research line focuses on improving the efficiency of a network is weight \emph{quantization}. The techniques in this area aim at reducing the number of bits necessary to represent the model weights: from the $32$ bits necessary to represent a float to only a few bits~\cite{quantized2016}. The state of the art network quantization techniques~\cite{2018alternating,ardakani2018learning} aims at quantizing the network weights using just $2$-$3$ bits per parameter. These approaches  proved effective on convolutional and recurrent neural networks.
Quantization strategies could be used in our proposed approach. However, to reduce the size of the term representations, we opt to instead focus on approaches to reduce the dimensionality of the term representations, and leave quantization of the stored embeddings to future work.

A third research line employed to speed-up neural networks is \emph{knowledge distillation}~\cite{DBLP:journals/corr/HintonVD15}. It aims to transform the knowledge embedded in a large network (called teacher) into a smaller network (called student). The student network is trained to reproduce the results of the teacher networks using a simpler network structure, with less parameters than those used in the teacher network. Several strategies have been proposed to distill knowledge in pretrained transformer networks such as BERT~\cite{Tang2019DistillingTK,Sanh2019DistilBERTAD,Jiao2019TinyBERTDB}.

Our \prettr~method is orthogonal to knowledge distillation of transformer network. In fact, our approach can be applied directly to any kind of transformer, including those produced by knowledge distillation.

\subsection{Neural Ranking Efficiency}\label{ssec:ranking}
Scalability and computational efficiency are central challenges in information retrieval. While the efficiency of learning to rank solutions for document re-ranking have been extensively studied~\cite{tois:quickscorer,tpds,fnt}, computational efficiency concerns have largely be ignored by prior work in neural ranking, prompting some to call for more attention to this matter~\cite{Hofsttter2019LetsMR}. That being said, some efforts do exist. \sm{For instance, \citet{zamani2018neural} investigate learning sparse query and document representations which allow for indexing.} \citet{ji-19} demonstrate that Locality-Sensitive Hashing (LSH) and other tricks can be employed to improve the performance of interaction-focused methods such as DRMM~\cite{guo-16}, KNRM~\cite{xiong-17}, and ConvKNRM~\cite{dai-18}. This approach does not work for transformer models, however, because further processing of the term embeddings is required (rather than only computing similarity scores between the query and document).

Within the realm of transformer-based models for ad-hoc ranking, to our knowledge only~\cite{MacAvaney2019CEDRCE} and~\cite{Nogueira2019DocumentEB} acknowledge that retrieval speed is substantially impacted by using a deep transformer network. As a result~\citet{Hofsttter2019LetsMR} call for more attention to be paid to run time. MacAvaney et al. find that limiting the depth of the transformer network can reduce the re-ranking time while yielding comparable ranking performance~\cite{MacAvaney2019CEDRCE}. Nogueira et al. find that their approach is faster than a transformer-based re-ranker, but it comes at a great cost to ranking performance: a trade-off that they state can be worthwhile in some situations~\cite{Nogueira2019DocumentEB}. In contrast with both these approaches, we employ \textit{part} of the transformer network at index time, and the remainder at query-time (for re-ranking). We find that this can yield performance on par with the full network, while significantly reducing the query time latency.

\section{Motivation}
\label{sec:motivation}

\begin{figure}[t]
\centering
\includegraphics[scale=0.58]{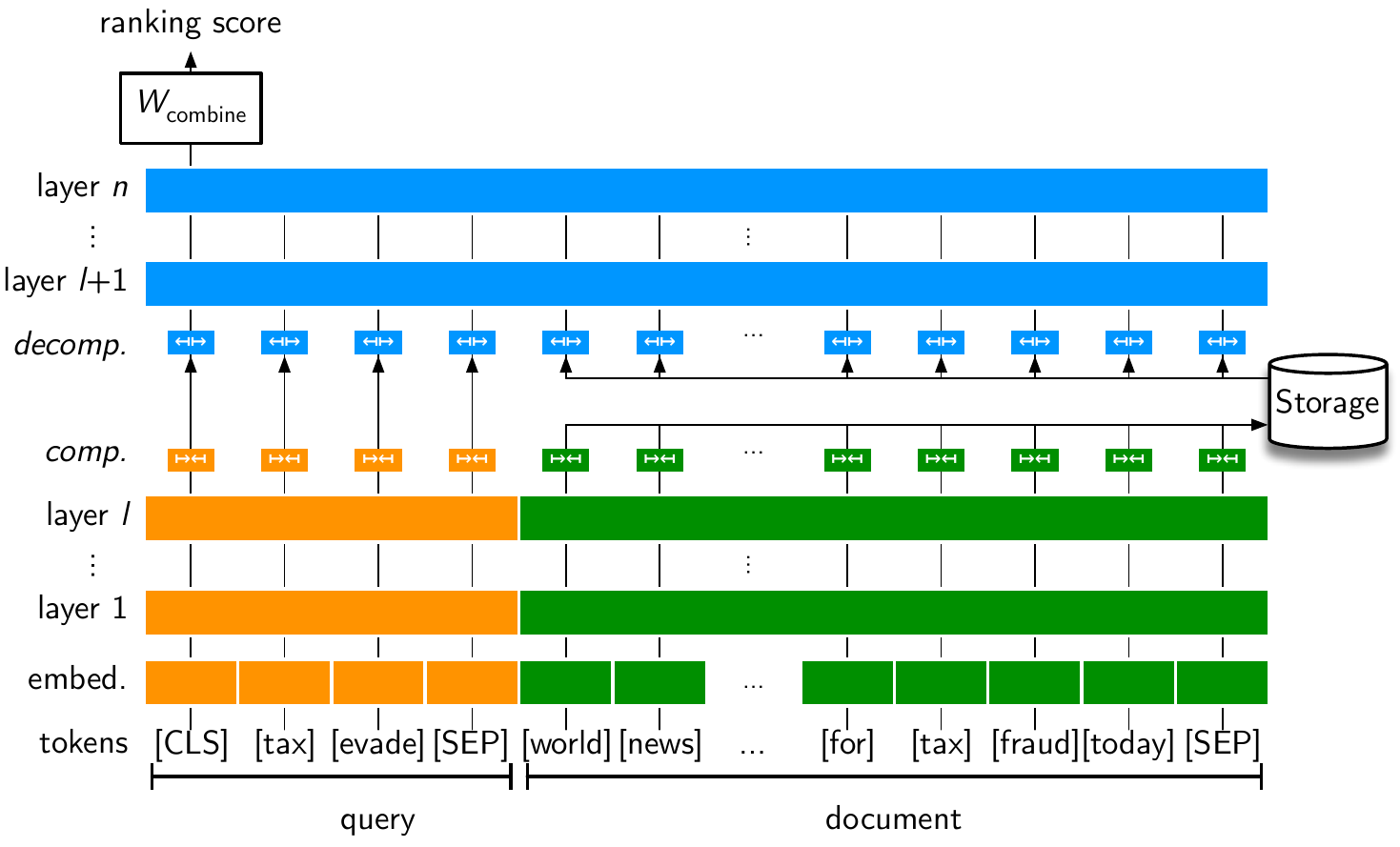}
\vspace{-1em}
\caption{Overview of \prettr{}. Compressed term representations for document layers 1 to $l$ are computed and stored at index time (green segments) while term representations for query layers 1 to $l$ (orange segments) and joint query-document representations for layers $l+1$ to $n$ (blue segments) are computed at query time to produce the final ranking score. Compression and decompression can optionally be applied between layers $l$ and $l+1$ to reduce the storage needed for the document term representations.
\label{fig:main_diagram}}
\vspace{-1em}
\end{figure}

\begin{table}[t!]
\centering
\caption{Table of symbols.}
\label{tab:symbols}
\small
{\renewcommand{\arraystretch}{0.8}
\begin{tabular}{cl}
\toprule
Symbol(s) & Definition \\
\midrule
$\query$ & Query \\
$\doc$ & Document \\
$R(\query,\doc)$ & Neural ranking architecture \\
$T(\mathbf{s})$ & Transformer network \\
$\mathbf{s}$ & a sequence of input tokens \\
$E$ & Embedding layer \\
$L_i$ & Transformer encoding layer \\
$\mathbf{s}_i$ & Transformer token representations after layer $i$ \\
$\mathbf{a}_i$ & Attention weights used in layer $i$ \\
$c$ & Classification representation \\
$d$ & Dimension of the classification representation \\
$m$ & Length of sequence $\mathbf{s}$ \\
$h$ & Number of attention heads per layer \\
$n$ & Number of layers in $T$ \\
$W_{combine}$ & Vanilla BERT weight combination \\
\multirow{2}{*}{$l$} & Layer number the transformer is executed for \\ 
                     & precomputing document term vectors \\
$e$ & Compressed size \\
$\textbf{r}$ & Compressed representation after layer $l$ \\
$W$/$\textbf{b}_{comp}$ & Compression parameters \\
$W$/$\textbf{b}_{decomp}$ & De-compression parameters \\
$\hat{\mathbf{s}}_{l}$ & De-compressed representation  after layer $l$\\
\bottomrule
\end{tabular}
}
\end{table}

Let a generic transformer network $T: \mathbf{s} \mapsto c$ map a sequence $\mathbf{s}$ of $m$ tokens (e.g., query and document terms) to a $d$-dimensional output representation $c\in\mathbb{R}^d$.
As depicted in Figure~\ref{fig:main_diagram}, the transformer network is composed by an initial embedding layer $E$ and by $n$ layers $L_1, \ldots, L_n$.
The embedding layer $E$ maps each of the $m$ input tokens into the initial $d$-dimensional token representations matrix $\mathbf{s}_0 \in \mathbb{R}^{m\times d}$. Each layer $L_i$ takes the token representations matrix $\mathbf{s}_{i-1}\in R^{m\times d}$ from the previous layer $L_{i-1}$ and produces a new representations matrix $\mathbf{s}_i\in \mathbb{R}^{m \times d}$. The specific representation used and operations performed in $E$ and $L_i$ depend on the specific transformer architecture (e.g., BERT uses token, segment, and position embeddings for the embedding layer $E$ and self-attention, a feed-forward layer, and batch normalization in each layer $L_i$). However, the primary and common component of each layer $L_i$ is the self-attention mechanism and associated procedure. When the transformer network is trained, every layer produces a self-attention tensor $\mathbf{a}_i\in \mathbb{R}^{h \times m \times m}$, where $h$ is the number of attention heads per layer, i.e., the number of attention ``representation subspaces'' per layer.
A general description of this process is given by Vaswani et al.~\cite{vaswani-17}, while different transformer architectures may have tweaks to this general structure or pre-training procedure.

We assume a special output classification token, e.g., \texttt{[CLS]} in BERT, is included as a token in $c$, and that the final representation of this token is used as the final output of the transformer network, i.e., $c = T(\mathbf{s})$.
Without loss of generality, here we only concern ourselves with the \texttt{[CLS]} output classification token, i.e., we ignore other token representation outputs; this is the special token representation that models such as BERT use to generate ranking scores.

We illustrate how neural transformer networks are used in a ranking scenario. We follow the Vanilla BERT model proposed by \citet{MacAvaney2019CEDRCE} and generalize it.
Let a ranking function $R(\query,\doc)\in\mathbb{R}$ map a query $\query$ and a document $\doc$ to a real-valued ranking score. 
Neural rankers based on transformer networks such as Vanilla BERT compute the ranking score by feeding the query-document pair into the transformer.
Given a query $\query$ and a document $\doc$, their tokens are concatenated into a suitable transformer input, e.g., $\mathbf{s} = \mathtt{[CLS]} ; \query ; \texttt{[SEP]} ; \doc ; \texttt{[SEP]}$, where ``;'' represents the concatenation operator.\footnote{We use the BERT convention of \texttt{[CLS]} and \texttt{[SEP]} to represent the classification and separation tokens, respectively.}
The output of the transformer network corresponding to this input is then linearly combined using a tuned weight matrix $W_{combine}\in\mathbb{R}^{d\times1}$ to compute the final ranking score as follows:

\begin{equation}
R(\query,\doc) = T\big(\mathtt{[CLS]} ; \query ; \texttt{[SEP]} ; \doc ; \texttt{[SEP]}\big)W_{combine}.
\end{equation}

The processing time of state-of-the-art neural rankers based on transformer networks is very high, e.g., approximately 50 documents ranked per second on a modern GPU, making such rankers impractical for most ad-hoc retrieval tasks. 

To gain an understanding of where are the most expensive components of a transformer network such as the Vanilla BERT model, we measure the run-times of the main steps of the model.
We find that most of the processing is performed in the computations involving the transformer's layers. In particular, about 50\% of the total time is spent performing attention-related tasks. 
Moreover, the feed-forward step of the transformer (consisting of intermediate and output in diagram) accounts for about 48\% of the total time, and is largely due to the large intermediate hidden representation size
for each token.
This breakdown motivates the investigation of possible solutions to reduce the processing time of transformer networks, in particular in reducing the time spent in traversing the transformer's layers.

\section{Proposed Solution}\label{sec:methodology}
We discuss how our \prettr{} approach improve the efficiency of processing queries using a transformer network by reducing the computational impact of the network's layers.

\subsection{PreTTR: Precomputing Transformer Term Representations}
We improve the query time performance of transformer models by precomputing document term representations partially through the transformer network (up to transformer layer $l$). We then use these representations at query time to complete the execution of the network when the query is known.

This is accomplished at model training time by applying an attention mask to layers $L_1,L_2,\ldots,L_{l}$, in which terms from the query are not permitted to attend to terms from the document and vice versa. In layers $L_{l+1},\ldots,L_{n}$, this attention mask is removed, permitting any token to attend to any other token.
Once trained, the model is used at both index and query time. At index time, documents are encoded (including the trailing \texttt{[SEP]} token)\footnote{There is evidence that the separator token performs an important function for pretrained transformer models, by acting as a no-op for the self-attention mechanism~\cite{clark-19}.} by the transformer model through layers $L_1,L_2,\ldots,L_{l}$ without a query present (Figure \ref{fig:main_diagram}, green segments). The token representations generated at index time at layer $L_l$ are then stored to be reused at query time (Figure \ref{fig:main_diagram}, document storage between layers $L_{l}$ and $L_{l+1}$). To answer a query, candidate documents are selected, e.g., the top  documents retrieved by a first-stage simple ranking model~\cite{fnt}, and precomputed term representations are loaded. The query terms (including the leading \texttt{[CLS]} and training \texttt{[SEP]} tokens) are encoded up to layer $L_l$ without a document present (Figure \ref{fig:main_diagram}, orange segments). Then, the representations from the query and the document are joined, and the remainder of the transformer network is executed over the entire sequence to produce a ranking score (Figure \ref{fig:main_diagram}, blue segments).

Since (1) the length of a query is typically much shorter than the length of a document, (2) the query representations can be re-used for each document being ranked, (3) each transformer layer takes about the same amount of time to execute, and (4) the time needed to perform term embedding is comparatively low, \prettr{} decreases by about $\frac{n-l}{n}$ the cost of traversing the transformer network layers. With a sufficiently large value of $l$, this results in considerable time savings. Note that this reduction can be at most equal to $\frac{1}{n}$ because, when $l=n$, no information about the document ever contributes to the ranking score, resulting in identical scores for every document. Moreover, we show experimentally that this can be further improved by limiting the computation of the final layer to only the \texttt{[CLS]} representation.

\subsection{Token Representation Compression}\label{sec:doctokcomp}

Although \prettr{} can reduce the run-time cost of traversing the first $l$ layers of the  transformer network at query time, the solution proposed might be costly in terms of storage requirements because the representation size $d$ is quite large (e.g., $1024$, $768$ or $512$ float values per token). To address this issue, we propose a new token compression technique that involves pre-training a simple encoder-decoder network. This network is able to considerably reduce the token representation size. We opt for this approach because it can fit seamlessly into the transformer network, while reducing the number of dimensions needed to represent each token. The compressor is added as an additional component of the transformer network between layers $L_l$ and $L_{l+1}$. We compress the input by using a simple feed-forward and normalization procedure, identical to the one used within a BERT layer to transform the output (but with a \textit{smaller} internal representation rather than a larger one). \sm{We optimize the weights for the compression network in two stages: (1) an initial pre-training stage on unlabeled data, and (2) a fine-tuning stage when optimizing for relevance.}

For a compressed size of $e$ values, a two-step procedure is used. First, the compressed representations $\textbf{r}\in\mathbb{R}^{m\times e}$ are built using $\textbf{r}=\textsc{gelu}(\textbf{s}_l W_{comp}+\textbf{b}_{comp})$, where $\textsc{gelu}(\cdot)$ is a Gaussian Error Linear Unit~\cite{hendrycks2016gaussian}, and $W_{comp}\in\mathbb{R}^{d\times e}$ and $\textbf{b}_{comp}\in\mathbb{R}^e$ are the new learned weight parameters. These compressed representations $\textbf{r}$ can be stored in place of $\textbf{s}_l$. Second, the compressed representations $\textbf{r}$ are then expanded back out to $\hat{\mathbf{s}}_{l}\in\mathbb{R}^{m\times d}$ via a second linear transformation involving the learned weight parameters $W_{decomp}$, $\mathbf{b}_{decomp}$, and batch normalization. The decompressed representations $\hat{\mathbf{s}_l}$ are then used in place of the original representation $\mathbf{s}_l$ for the remaining layers of the transformer.

In preliminary experiments, we found the compression and decompression parameters to be difficult to learn jointly with the ranker itself. Thus, we instead propose a pre-training approach to provide an effective initialization of these parameters. We want the transformer network with the compression mechanism to behave similarly to that of the network without such compression: we do not necessarily care about the exact representations themselves. Thus, we use an attention-based loss function. More specifically, we optimize our compression/decompression network to reduce the mean squared error of the attention scores in the last $n-l$ layers of the compressed transformer network and the original transformer network. 
Thus, the loss function we use to train our compression and decompression network is:
\begin{equation}\label{eq:attnloss}\small
\mathcal{L}(\mathbf{a}_{l+1},\ldots,\mathbf{a}_n,\hat{\mathbf{a}}_{l+1},\ldots,\hat{\mathbf{a}}_n)=\frac{1}{n-l}\sum_{i=l+1}^{n}\textsc{MSE}(\mathbf{a}_i,\hat{\mathbf{a}}_i),
\end{equation}
where $a_i$ represents the attention scores at layer $i$ from the unmodified transformer network, $\hat{a_i}$ represents the attention scores at layer $i$ from the transformer network with the compression unit, and $\textsc{MSE}(\cdot)$ is the mean squared error function. With this loss function, the weights can be \sm{pre-}trained on a massive amount of unlabeled text. \sm{We use this procedure as an initial pre-training step; we further fine-tune the weights when optimizing the entire ranking network for relevance.}

\section{Experimental Setup}\label{sec:experiments}
We detail the setup employed in our experiments: the datasets, namely TREC WebTrack 2012 and TREC Robust 2004, and the transformer networks we use, i.e., Vanilla BERT and some of its variants. Then, we discuss the training procedure adopted in training the transformer networks and our proposed compression/decompression technique. Details about the evaluation metrics and the baselines used conclude the section.

\subsection{Datasets}
We test \prettr{} on two datasets, namely TREC WebTrack 2012 and TREC Robust 2004. Table~\ref{tab:datasets} summarizes some salient statistics about the two datasets.

\begin{table}[h]
\small
\centering
\caption{Datasets characteristics.\label{tab:datasets}}
\vspace{-1em}
\begin{tabular}{lcc}
\toprule
 & WebTrack 2012 & Robust 2004 \\
\cmidrule(lr){2-2} \cmidrule(lr){3-3}
Domain & Web & Newswire \\
Document collection & ClueWeb09-B & TREC Disks 4 \& 5 \\
\# Queries & 50 & 249 \\
\# Documents & $50M$ & $528k$ \\
Tokens / query & 2.0 & 2.7 \\
Judgments / query & 321 & $1.2k$ \\
\bottomrule
\end{tabular}
\vspace{-1em}
\end{table}

The TREC WebTrack 2012 dataset consists of web queries and relevance judgments from the ClueWeb09-B document collection. We use relevance judgments from 2012 for test and the ones from 2011 for validation. The relevance judgments available from the remaining years of the TREC WebTrack, i.e., 2009, 2010, 2013, and 2014 are used for training. Note that, while the TREC WebTrack 2009--12 have been evaluated on the ClueWeb09-B document collection, the TREC WebTrack 2013--14 have been evaluated on the ClueWeb12~\cite{Hui2017PACRRAP} document collection.\footnote{\url{https://lemurproject.org/clueweb09/} and \url{https://lemurproject.org/clueweb12/}.} We generate the training samples by using the corresponding document collection. This is the setup used by several other works on TREC WebTrack 2012, e.g., \cite{Hui2017PACRRAP,MacAvaney2019CEDRCE}.

TREC Robust 2004 consists of 249 news queries. For these experiments, we use a standard $k$-fold evaluation ($k=5$) where each iteration uses three folds for training, one for validation, and a final held-out fold for testing. We perform this evaluation by using the five folds provided by Huston and Croft~\cite{huston2014parameters}.

\subsection{Transformer Networks}
We use the Vanilla transformer model from \cite{MacAvaney2019CEDRCE}. This model yields comparable performance to other leading formulations, while being simpler, e.g., no paragraph segmentation required, as is needed by \texttt{FirstP}/\texttt{MaxP}/\texttt{SumP}~\cite{Dai2019DeeperTU}, or alternative training datasets and sentence segmentation, as required \sm{by the system of~\citet{Yang2019SimpleAO}}. Vanilla BERT encodes as much of the document as possible (adhering to the transformer maximum input length constraint), and averages the classification embeddings when multiple document segments are required. We employ the same optimal hyper-parameters for the model presented in~\cite{MacAvaney2019CEDRCE}. For our primary experiments, we use the pretrained \texttt{bert-base-uncased}~\cite{devlin-19}. We do not test with the \texttt{large} variants of BERT because the larger model exhibits only marginal gains for ranking tasks, while being considerably more expensive to run~\cite{Nogueira2019PassageRW}. To show the generality of our approach we present tests conducted also for other pretrained transformers in Section~\ref{sec:other_transformers}: a version of BERT that was more effectively pre-trained, i.e., RoBERTa~\cite{Liu2019RoBERTaAR} (\texttt{roberta-base}) and a smaller (distilled) version of BERT, i.e.,  DistilBERT~\cite{Sanh2019DistilBERTAD} (\texttt{distilbert-base-uncased}).

\subsection{Training}
We train all transformer models using pairwise softmax loss~\cite{Dehghani2017NeuralRM} and the Adam optimizer~\cite{Kingma2015AdamAM} with a learning rate of $2\times10^{-5}$. We employ a batch size of $16$ pairs of relevant and non-relevant documents with gradient accumulation. Training pairs are selected randomly from the top-ranked documents in the training set, where documents that are labeled as relevant are treated as positive, and other top-ranked documents are considered negative. Every $32$ batches, the model is validated, and the model yielding the highest performance on the validation set is selected for final evaluation. 

For training the document term compressor/decompressor (as described in Section~\ref{sec:doctokcomp}), we use the Wikipedia text from the TREC Complex Answer Retrieval (CAR) dataset~\cite{Dietz2017} (version 2.0 release). This dataset was chosen because it overlaps with the data on which BERT was originally trained on, i.e., Wikipedia, and was used both for evaluation of passage ranking approaches~\cite{Nanni2017BenchmarkFC} and as a weak supervision dataset for training neural models~\cite{MacAvaney2019ContentBasedWS}. We sample text pairs using combinations of headings and paragraphs. Half the pairs use the heading associated with the paragraph, and the other half use a random heading from a different article, akin to the next sentence classification used in BERT pre-training. The compression and decompression parameters ($W_{comp}$, $\textbf{b}_{comp}$, $W_{decomp}$, and $\textbf{b}_{decomp}$) are trained to minimize the difference in attention scores, as formulated in Eq.~\eqref{eq:attnloss}. We found that the compressor training process converged by $2M$ samples.

\subsection{Evaluation}
Since the transformer network is employed as a final-stage re-ranker, we evaluate the performance of our approach on each dataset using two precision-oriented metrics. Our primary metric for both datasets is P@20 (also used for model validation). Following the evaluation convention from prior work~\cite{MacAvaney2019CEDRCE}, we use ERR@20 for TREC WebTrack 2012 and nDCG@20 for TREC Robust 2004 as secondary metrics.

We also evaluate the query-time latency of the models. We conduct these experiments using commodity hardware: one GeForce GTX 1080 Ti GPU. To control for factors such as disk latency, we assume the model and term representations are already loaded in the main memory. In other words, we focus on the impact of the model computation itself. However, the time spent moving the data to and from the GPU memory is included in the time.

\subsection{Baselines}
The focus of this work is to reduce the query-time latency of using Vanilla transformer models, which are among the state-of-the-art neural ranking approaches. Thus, our primary baseline is the unmodified Vanilla transformer network. To put the results in context, we also include the BM25 results tuned on the same training data. We tune BM25 using grid search with Anserini's implementation~\cite{Yang2017AnseriniET}, over $k_1$ in the range of 0.1--4.0 (by 0.1) and $b$ in the range of 0.1--1.0 (by 0.1). We also report results for CEDR-KNRM~\cite{MacAvaney2019CEDRCE}, which outperform the Vanilla transformer approaches. However, it come with its own query-time challenges. Specifically, since it uses the term representations from every layer of the transformer, this would require considerably more storage. To keep our focus on the typical approach, i.e., using the \texttt{[CLS]} representation for ranking, we leave it to future work to investigate ways in which to optimize the CEDR model.\footnote{\edit{We note that techniques such as LSH hashing can reduce the storage requirements for CEDR, as it uses the representations to compute query-document similarity matrices, as demonstrated by~\cite{ji-19}.}} We also report results for Birch~\cite{Yilmaz2019ApplyingBT}, which exploits transfer learning from the TREC Microblog dataset. To keep the focus of this work on the effect of pre-computation, we opt to evaluate in the single-domain setting.

\section{Results and Discussion}
\label{sec:analysis}
We report the results of a comprehensive experimental evaluation of the proposed \prettr{} approach. In particular, we aim at investigating the following research questions:
\begin{itemize}
\item[RQ1] What is the impact of \prettr{} on the effectiveness of the Vanilla BERT transformer network in ad-hoc ranking? (Section~\ref{sec:results_prettr})
\item[RQ2] What is the impact of the token representation compression on the effectiveness of \prettr{}? (Section~\ref{sec:results_comp})
\item[RQ3] What is the impact of the proposed \prettr{} approach on the efficiency of Vanilla BERT when deployed as a second stage re-ranker? (Section~\ref{sec:results_latency})
\item[RQ4] What is the impact of \prettr{} when applied to first $n-1$ layers of a transformer network? (Section~\ref{sec:resutls_singlelayer})
\item[RQ5] What is the impact of \prettr{} when applied to different transformer networks such as RoBERTA and DistilBERT? (Section~\ref{sec:other_transformers})
\end{itemize}

\subsection{Precomputing Transformer Term Representations}\label{sec:results_prettr}

To answer RQ1 we first evaluate the effect of the precomputation of term representations. Table~\ref{tab:part-sep} provides a summary of the ranking performance of \prettr{}-based Vanilla BERT at layer $l$. \edit{At lower values of $l$, the ranking effectiveness remains relatively stable, despite some minor fluctuations. We note that these fluctuations are not statistically significant when compared with the base model (paired t-test, 99\% confidence interval) and remain considerably higher than the tuned BM25 model. We also tested using a two one-sided equivalence (TOST) and found similar trends (i.e., typically the the significant differences did not exhibit significant equivalence.)}
In the case of TREC WebTrack 2012, the model achieves comparable P@20 performance w.r.t. the base model with only a single transformer layer ($12$), while the first $11$ layers are precomputed. Interestingly, the ERR@20 suffers more than P@20 as more layers are precomputed. This suggests that the model is able to identify generally-relevant documents very effectively with only a few transformer layers, but more are required to be able to identify the subtleties that contribute to greater or lesser degrees of relevance. \edit{Although it would ideally be best to have comparable ERR@20 performance in addition to P@20, the substantial improvements that this approach offers in terms of query-time latency (see Section~\ref{sec:results_latency}) may make the trade-off worth it, depending on the needs of the application.}

On the TREC Robust 2004 newswire collection, precomputing the first $10$ layers yields comparable P@20 performance w.r.t. the base model. Interestingly, although $l=11$ yields a relatively effective model for WebTrack, Robust performance significantly suffers in this setting, falling well below the BM25 baseline. \edit{We also observe a significant drop in nDCG@20 performance at $l=8$, while P@20 performance remains stable until $l=11$. This is similar to the behavior observed on WebTrack: as more layers are precomputed, the model has a more difficult time distinguishing graded relevance.}

\begin{table}
\centering
\caption{Breakdown of ranking performance when using a \prettr{}-based Vanilla BERT ranking, joining the encodings at layer $l$. Statistically significant differences with the base model are indicated by $\downarrow$ (paired t-test by query, $p<0.01$).
}
\vspace{-1em}
\label{tab:part-sep}

\begin{tabular}{lrrrr}
\toprule
& \multicolumn{2}{c}{WebTrack 2012} & \multicolumn{2}{c}{Robust 2004} \\
\cmidrule(lr){2-3} \cmidrule(lr){4-5}
Ranker & P@20 & ERR@20 & P@20 & nDCG@20 \\
\midrule

Base & \bf0.3460 & 0.2767 & 0.3784 & 0.4357 \\
$l=1$ & 0.3270 & \bf0.2831 & 0.3851 & \bf0.4401 \\
$l=2$ & 0.3170 & 0.2497 & 0.3821 & 0.4374 \\
$l=3$ & 0.3440 & 0.2268 & \bf0.3859 & 0.4386 \\
$l=4$ & 0.3280 & 0.2399 & 0.3701 & 0.4212 \\
$l=5$ & 0.3180 & 0.2170 & 0.3731 & 0.4214 \\
$l=6$ & 0.3270 & 0.2563 & 0.3663 & 0.4156 \\
$l=7$ & 0.3180 & 0.2255 & 0.3656 & 0.4139 \\
$l=8$ & 0.3140 & 0.2344 & 0.3636 & $\downarrow$ 0.4123 \\
$l=9$ & 0.3130 & 0.2297 & 0.3644 & $\downarrow$ 0.4106 \\
$l=10$ & 0.3360 & 0.2295 & 0.3579 & $\downarrow$ 0.4039 \\
$l=11$ & 0.3380 & $\downarrow$ 0.1940 & $\downarrow$ 0.2534 & $\downarrow$ 0.2590 \\

\midrule
Tuned BM25 & 0.2370 & 0.1418 & 0.3123 & 0.4140 \\
\midrule
Vanilla BERT~\cite{MacAvaney2019CEDRCE} & - & - & 0.4042 & 0.4541 \\
CEDR-KNRM~\cite{MacAvaney2019CEDRCE} & - & - & 0.4667 & 0.5381 \\
Birch~\cite{Yilmaz2019ApplyingBT} & - & - & 0.4669 & 0.5325 \\
\bottomrule
\end{tabular}
\vspace{-1em}
\end{table}

\begin{table*}
\centering
\caption{Ranking performance at various compression sizes. Statistically significant increases and decreases in ranking performance (compared to the model without compression) are indicated with $\uparrow$ and $\downarrow$, respectively (paired t-test by query, $p<0.01$). \edit{We mark columns with * to indicate cases in which the uncompressed model (none) significantly underperforms the Base model performance (from Table~\ref{tab:part-sep}).}}
\label{tab:compression}
\vspace{-1em}
{\renewcommand{\arraystretch}{0.85}
\begin{tabular}{l@{\hskip 0.3in}rrrrrrrrrr}
\multicolumn{11}{c}{TREC WebTrack 2012} \\
\toprule
&\multicolumn{5}{c}{P@20}&\multicolumn{5}{c}{ERR@20} \\\cmidrule(lr){2-6}\cmidrule(lr){7-11}
Compression & $l=7$ & $l=8$ & $l=9$ & $l=10$ & $l=11$ & $l=7$ & $l=8$ & $l=9$ & $l=10$ & * $l=11$ \\
\midrule

(none) & 0.3180 & 0.3140 & 0.3130 & \bf0.3360 & \bf0.3380 & 0.2255 & \bf0.2344 & 0.2297 & \bf0.2295 & 0.1940 \\
$e=384$ (50\%) & \bf0.3430 & \bf0.3260 & 0.2980 & \bf0.3360 & 0.3090 & 0.2086 & 0.2338 & 0.1685 & 0.2233 & \bf0.2231 \\
$e=256$ (67\%) & 0.3380 & 0.3120 & \bf$\uparrow$ 0.3440 & 0.3260 & 0.3250 & \bf$\uparrow$ 0.2716 & 0.2034 & \bf$\uparrow$ 0.2918 & 0.1909 & 0.2189 \\
$e=128$ (83\%) & 0.3100 & 0.3210 & 0.3320 & 0.3220 & 0.3370 & 0.2114 & 0.2234 & 0.2519 & 0.2239 & 0.2130 \\

\bottomrule
\end{tabular}

\vspace{4mm}

\begin{tabular}{l@{\hskip 0.3in}rrrrrrrrrr}
\multicolumn{11}{c}{TREC Robust 2004} \\
\toprule
&\multicolumn{5}{c}{P@20}&\multicolumn{5}{c}{nDCG@20} \\\cmidrule(lr){2-6}\cmidrule(lr){7-11}
Compression & $l=7$ & $l=8$ & $l=9$ & $l=10$ & * $l=11$ & $l=7$ & * $l=8$ & * $l=9$ & * $l=10$ & * $l=11$ \\
\midrule

(none) & \bf0.3656 & \bf0.3636 & \bf0.3644 & \bf0.3579 & 0.2534 & \bf0.4139 & \bf0.4123 & \bf0.4106 & \bf0.4039 & 0.2590 \\
$e=384$ (50\%) & 0.3587 & $\downarrow$ 0.3369 & $\downarrow$ 0.3435 & 0.3522 & \bf0.2687 & 0.4098 & $\downarrow$ 0.3720 & $\downarrow$ 0.3812 & 0.3895 & \bf$\uparrow$ 0.2807 \\
$e=256$ (67\%) & $\downarrow$ 0.2950 & 0.3623 & $\downarrow$ 0.2695 & 0.3535 & 0.2635 & $\downarrow$ 0.3130 & 0.4074 & $\downarrow$ 0.2753 & 0.3983 & 0.2694 \\
$e=128$ (83\%) & $\downarrow$ 0.2461 & $\downarrow$ 0.2530 & $\downarrow$ 0.2499 & $\downarrow$ 0.2607 & 0.2655 & $\downarrow$ 0.2454 & $\downarrow$ 0.2568 & $\downarrow$ 0.2533 & $\downarrow$ 0.2608 & 0.2713 \\

\bottomrule
\end{tabular}
}
\vspace{-0.5em}
\end{table*}

We observe that the highest-performing models (metric in bold) are not always the base model. \sm{However, we note that these scores do not exhibit statistically significant differences when compared to the base model.}

In summary, we answer RQ1 by showing that Vanilla BERT can be successfully trained by limiting the interaction between query terms and document terms, and that this can have only a minimal impact on ranking effectiveness, \edit{particularly in terms in the precision of top-ranked documents}. This is an important result because it shows that document term representations can be built independently of the query at index time.

\subsection{Term Representation Compression}\label{sec:results_comp}

To answer RQ2, we run the Vanilla BERT model with varying sizes $e$ of the compressed embedding representations over the combination layers $l$ that give the most benefit to query latency time (i.e., $l = 7, 8, 9, 10, 11$). \sm{Layers $l\leq6$ are not considered because they provide less computational benefit (taking about one second or more per 100 documents, see Section~\ref{sec:results_latency}).} See Table~\ref{tab:compression} for a summary of the results on TREC WebTrack 2012 and Robust 2004. We find that the representations can usually be compressed down to at least $e=256$ (67\% of the original dimension of 768) without substantial loss in ranking effectiveness. In Robust, we observe a sharp drop in performance at $e=128$ (83\% dimension compression) at layers 7--10. There is no clear pattern for which compression size is most effective for WebTrack 2012. \sm{Note that these differences are generally not statistically significant.} This table shows that, to a point, there is a trade-off between the size of the stored representations and the effectiveness of the ranker.

Without any intervention, approximately 112TB of storage would be required to store the full term vectors for ClueWeb09-B (the document collection for TREC WebTrack 2012). For web collections, this can be substantially reduced by eliminating undesirable pages, such as spam. Using recommended settings for the spam filtering approach proposed by \citet{Cormack2010EfficientAE} for ClueWeb09-B, the size can be reduced to about 34TB. Using our compression/decompression approach, the storage needed can be further reduced, depending on the trade-off of storage, query-time latency, and storage requirements. If using a dimension $e=128$ for the compressed representation (with no statistically significant differences in effectiveness on WebTrack), the size is further reduced to 5.7TB, which yields a 95\% of space reduction. We also observed that there is little performance impact by using 16-bit floating point representations, which further reduces the space to about 2.8TB. Although this is still a tall order, it is only about 2.5\% of the original size, and in the realm of reasonable possibilities. We leave it to future work to investigate further compression techniques, such as kernel density estimation-based quantization~\cite{Seo2019EfficientWQ}.

Since the size scales with the number of documents, the storage requirements are far less for smaller document collections such as newswire. Document representations for the TREC Disks 4 \& 5 (the document collection for the Robust 2004) can be stored in about 195GB, without any filtering and using the more effective $e=256$  for the dimension of the compressed representation.

In summary, regarding RQ2, we \fm{show} that, through our compression technique, one can reduce the storage requirements of \prettr{}. With a well-trained compression and decompression weights, this can have minimal impact on ranking effectiveness.

\begin{table}
\centering
\small
\caption{Vanilla BERT query-time latency measurements for re-ranking the top 100 documents on TREC WebTrack 2012 and TREC Robust 2004. The latency is broken down into time to compute query representations up through layer $l$, the time to decompress document term representations, and the time to combine the query and document representations from layer $l+1$ to layer $n$. The $l=11$ setting yields a 42$\times$ speedup for TREC WebTrack, while \edit{not significantly reducing the ranking performance}.}
\vspace{-1em}
\label{tab:speed}
\scalebox{0.92}{
\begin{tabular}{lrrrrrr}
\toprule
 & \multicolumn{5}{c}{TREC WebTrack 2012} & \multicolumn{1}{c}{Robust04} \\
\cmidrule(lr){2-6} \cmidrule(lr){7-7}
Ranker & Total & Speedup & Query & Decom. & Combine & Total \\
\midrule

Base & 1.941s & (1.0$\times$) & -  & - & - & 2.437s \\
$l=1$ & 1.768s & (1.1$\times$) &\bf2ms  & 10ms & 1.756s & 2.222s \\
$l=2$ & 1.598s & (1.2$\times$) & 3ms  & 10ms & 1.585s & 2.008s \\
$l=3$ & 1.423s & (1.4$\times$) & 5ms  & 10ms & 1.409s & 1.792s \\
$l=4$ & 1.253s & (1.5$\times$) & 6ms  & 10ms & 1.238s & 1.575s \\
$l=5$ & 1.080s & (1.8$\times$) & 7ms  & 10ms & 1.063s & 1.356s \\
$l=6$ & 0.906s & (2.1$\times$) & 9ms  & 10ms & 0.887s & 1.138s \\
$l=7$ & 0.735s & (2.6$\times$) & 10ms  & 10ms & 0.715s  & 0.922s\\ 
$l=8$ & 0.562s & (3.5$\times$) & 11ms  & 10ms & 0.541s & 0.704s \\
$l=9$ & 0.391s & (5.0$\times$) & 12ms  & 10ms & 0.368s & 0.479s \\
$l=10$ & 0.218s & (8.9$\times$) & 14ms  & 10ms & 0.194s & 0.266s \\
$l=11$ &\bf0.046s &\bf(42.2$\times$) & 15ms  & 10ms &\bf0.021s & \bf0.053s \\

\bottomrule
\end{tabular}
}
\vspace{-1em}
\end{table}

\subsection{Re-ranking Efficiency}\label{sec:results_latency}

The reduction of the re-ranking latency achieved by our proposed \prettr{} is considerable. To answer RQ3, in Table~\ref{tab:speed} we report an analysis of the re-ranking latency of \prettr{}-based Vanilla BERT when precomputing the token representations at a specific layer $l$ and a comparison against the base model, i.e., Vanilla BERT. Without our approach, re-ranking the top $100$ results for a query using Vanilla BERT takes around $2$ seconds.
Instead, when using \prettr{}-based Vanilla BERT at layer $l=11$, which yields \sm{comparable P@20 performance} to the base model on the TREC WebTrack 2012 collection, the re-ranking process takes $46$ milliseconds for $100$ documents, i.e., we achieve a $42.0\times$ speedup. One reason this performance is achievable is because the final layer of the transformer network does not need to compute the representations for each token; only the representations for the \texttt{[CLS]} token are needed, since it is the only token used to compute the final ranking score. Thus, the calculation of a full self-attention matrix is not required. Since the \texttt{[CLS]} representation is built in conjunction with the query, it alone can contain a summary of the query terms. Furthermore, since the query representation in the first $l$ layers is independent of the document, these representations are re-used among all the documents that are re-ranked. Of the time spent during re-ranking for $l=11$, 32\% of the time is spent building the query term representation, 21\% of the time is spent decompressing the document term representations, and the remainder of the time is spent combining the query and document representations. Moreover, when using \prettr{}-based Vanilla BERT at layer $l=10$, the transformer network needs to perform a round of computations on all the term representations. Nevertheless, in this case, our \prettr{} approach leads to a substantial speedup of $8.9\times$ w.r.t. Vanilla BERT. We also observe that the time to decompress the term representations (with $e=256$) remains a constant overhead, as expected. We observe a similar trend when timing the performance of Robust 2004\edit{, though we would recommend using $l\leq10$ for this dataset, as $l=11$ performs poorly in terms of ranking effectiveness. Nonetheless, at $l=10$, Robust achieves a $9.2\times$ speedup, as compared to the full model.}

In summary, regarding RQ3, we show that the \prettr{} approach can save a considerable amount of time at query-time, as compared to the full Vanilla BERT model. These time savings can make it practical to run transformer-based rankers in a real-time query environment.

\subsection{Single Layer Ranking ($l=11$)}\label{sec:resutls_singlelayer}
We answer RQ4 by highlighting a first interesting difference between the WebTrack and the Robust ranking performance: the effectiveness at $l=11$ (Table~\ref{tab:part-sep}). \sm{For WebTrack, the performance is comparable in terms of P@20, but suffers in terms of ERR@20.} For Robust, the performance suffers drastically. We attribute this to differences in the dataset characteristics. First, let us consider what happens in the $l=11$ case. Since it is the final layer and only the representation of the \texttt{[CLS]} token is used for ranking, the only attention comparisons that matter are between the \texttt{[CLS]} token and every other token (not a full comparison between every pair of tokens, as is done in other layers). Thus, a representation of the entire query must be stored in the \texttt{[CLS]} representation from layer $11$ to provide an effective comparison with the remainder of the document, which will have no contribution from the query. Furthermore, document token representations will need to have their context be fully captured in a way that is effective for the matching of the \texttt{[CLS]} representation. Interestingly, this setting blurs the line between representation-focused and interaction-focused neural models.

Now we will consider the characteristics of each dataset. From Table~\ref{tab:datasets}, we find that the queries in the TREC WebTrack 2012 are typically shorter (mean: 2.0, median: 2, stdev: 0.8) than those from Robust (mean: 2.7, median: 3, stdev: 0.7). This results in queries that are more qualified, and may be more difficult to successfully represent in a single vector.

To answer RQ4, we observe that the ranking effectiveness when combining with only a single transformer layer can vary depending on dataset characteristics. We find that in web collections (an environment where query-time latency is very important), it may be practical to use \prettr{} in this way while maintaining high precision of the top-ranked documents.

\subsection{\prettr{} for Other Transformers}
\label{sec:other_transformers}
Numerous pre-trained transformer architectures exist. We now answer RQ5 by showing that \prettr{} is not only effective on BERT, but its ability of reducing ranking latency by preserving quality holds also on other transformer variants. We investigate both the popular RoBERTa~\cite{Liu2019RoBERTaAR} model and the DistilBERT~\cite{Sanh2019DistilBERTAD} model. These represent a model that uses a more effective pre-training process, and a smaller network size (via model distillation), respectively. Results for this experiment are shown in Table~\ref{tab:other_transformers}. We first observe that the unmodified RoBERTa model performs comparably with the BERT model, while the DistilBERT model performs slightly worse. This suggests that model distillation alone may not be a suitable solution to address the poor query-time ranking latency of transformer networks. With each value of $l$, we observe similar behavior to BERT: P@20 remains relatively stable, while ERR@20 tends to degrade. Interestingly, at $l=2$ DistilBERT's ERR@20 performance peaks at 0.2771. However, this difference is not statistically significant, and thus we cannot assume it is not due to noise.

We tested the query-time latency of RoBERTa and DistilBERT in the same manner as described in Section~\ref{sec:results_latency}. With 12 layers and a similar neural architecture, RoBERTa exhibited similar speedups as BERT, with up to a $56.3\times$ speedup at $l=11$ (0.041s per 100 documents, down from 1.89s). With only 6 layers, the base DistilBERT model was faster (0.937s), and was able to achieve a speedup of $24.1\times$ with $l=5$ (0.035s).

In summary, we show that the \prettr{} approach can be successfully generalized to other transformer networks (RQ5). We observed similar trends to those we observed with BERT in two transformer variants, both in terms of ranking effectiveness and efficiency.

\begin{table}
\centering
\caption{WebTrack 2012 using two other Vanilla transformer architectures: RoBERTa and DistilBERT. Note that DistilBERT only has 6 layers; thus we only evaluate $l\in[1,5]$ for this model. There are no statistically significant differences between the Base Model and any of the \prettr{} variants (paired t-test, $p<0.01$).}
\vspace{-1em}
\label{tab:other_transformers}
{\renewcommand{\arraystretch}{1.0}
\begin{tabular}{lrrrr}
\toprule
& \multicolumn{2}{c}{RoBERTA~\cite{Liu2019RoBERTaAR}} & \multicolumn{2}{c}{DistilBERT~\cite{Sanh2019DistilBERTAD}} \\
\cmidrule(lr){2-3} \cmidrule(lr){4-5}
Ranker & P@20 & ERR@20 & P@20 & ERR@20 \\
\midrule

Base & 0.3370 & 0.2609 & 0.3110 & 0.2293 \\
$l=1$ & 0.3380 & \bf0.2796 & 0.3220 & 0.1989 \\
$l=2$ & 0.3370 & 0.2207 & 0.3340 & \bf0.2771 \\
$l=3$ & 0.3530 & 0.2669 & 0.3070 & 0.1946 \\
$l=4$ & \bf0.3620 & 0.2647 & 0.3350 & 0.2281 \\
$l=5$ & 0.2950 & 0.1707 & \bf0.3350 & 0.2074 \\
$l=6$ & 0.3000 & 0.1928 & - & - \\
$l=7$ & 0.3350 & 0.2130 & - & - \\
$l=8$ & 0.3220 & 0.2460 & - & - \\
$l=9$ & 0.3180 & 0.2256 & - & - \\
$l=10$ & 0.3140 & 0.1603 & - & - \\
$l=11$ & 0.3210 & 0.2241 & - & - \\

\bottomrule
\end{tabular}
}
\vspace{-1em}
\end{table}

\section{Conclusions and Future Work}
\label{sec:conclusions}

Transformer networks, such as BERT, present a considerable opportunity to improve ranking effectiveness~\cite{Nogueira2019PassageRW,Dai2019DeeperTU,MacAvaney2019CEDRCE}. However, relatively  little attention has been paid to the effect that these approaches have on query execution time. In this work, we showed that these networks can be trained in a way that is more suitable for query-time latency demands. Specifically, we showed that web query execution time can be improved by up to $42\times$ \edit{for web document ranking}, with minimal impact on P@20. Although this approach requires storing term representations for documents in the collection, we proposed an approach to reduce this storage required by 97.5\% by pre-training a compression/decompression function and using reduced-precision (16 bits) floating point arithmetic. We experimentally showed that the approach works across transformer architectures, and we demonstrated its effectiveness on both web and news search. These findings are particularly important for large-scale search settings, such as web search, where query-time latency is critical.

This work is orthogonal to other efforts to reign in the execution time of transformer networks.
There are challenges related to the application of more advanced networks, such as CEDR~\cite{MacAvaney2019CEDRCE}, which require the computation or storage of additional term representations. 
Future work could investigate how approaches like LSH-hashing~\cite{ji-19} could be used to help accomplish this.
Furthermore, our observation that comparable ranking performance can be achieved using a compression layer raises questions about the importance of the feed-forward step in each transformer layer.

\section*{Acknowledgments}
Work partially supported by the ARCS Foundation. Work partially supported by the Italian Ministry of Education and Research (MIUR) in the framework of the CrossLab project (Departments of Excellence). Work partially supported by the BIGDATAGRAPES project funded by the EU Horizon 2020 research and innovation programme under grant agreement No. 780751, and by the OK-INSAID project funded by the Italian Ministry of Education and Research (MIUR) under grant agreement No. ARS01\_00917.

\bibliographystyle{ACM-Reference-Format}
\balance
\bibliography{biblio}


\begin{thebibliography}{50}


\ifx \showCODEN    \undefined \def \showCODEN     #1{\unskip}     \fi
\ifx \showDOI      \undefined \def \showDOI       #1{#1}\fi
\ifx \showISBNx    \undefined \def \showISBNx     #1{\unskip}     \fi
\ifx \showISBNxiii \undefined \def \showISBNxiii  #1{\unskip}     \fi
\ifx \showISSN     \undefined \def \showISSN      #1{\unskip}     \fi
\ifx \showLCCN     \undefined \def \showLCCN      #1{\unskip}     \fi
\ifx \shownote     \undefined \def \shownote      #1{#1}          \fi
\ifx \showarticletitle \undefined \def \showarticletitle #1{#1}   \fi
\ifx \showURL      \undefined \def \showURL       {\relax}        \fi
\providecommand\bibfield[2]{#2}
\providecommand\bibinfo[2]{#2}
\providecommand\natexlab[1]{#1}
\providecommand\showeprint[2][]{arXiv:#2}

\bibitem[\protect\citeauthoryear{Ardakani, Ji, Smithson, Meyer, and
  Gross}{Ardakani et~al\mbox{.}}{2019}]%
        {ardakani2018learning}
\bibfield{author}{\bibinfo{person}{Arash Ardakani}, \bibinfo{person}{Zhengyun
  Ji}, \bibinfo{person}{Sean~C Smithson}, \bibinfo{person}{Brett~H Meyer},
  {and} \bibinfo{person}{Warren~J Gross}.} \bibinfo{year}{2019}\natexlab{}.
\newblock \showarticletitle{Learning Recurrent Binary/Ternary Weights}. In
  \bibinfo{booktitle}{\emph{{International Conference on Learning
  Representations}}}.
\newblock
\urldef\tempurl%
\url{http://arxiv.org/abs/1809.11086}
\showURL{%
\tempurl}


\bibitem[\protect\citeauthoryear{Clark, Khandelwal, Levy, and Manning}{Clark
  et~al\mbox{.}}{2019}]%
        {clark-19}
\bibfield{author}{\bibinfo{person}{Kevin Clark}, \bibinfo{person}{Urvashi
  Khandelwal}, \bibinfo{person}{Omer Levy}, {and}
  \bibinfo{person}{Christopher~D. Manning}.} \bibinfo{year}{2019}\natexlab{}.
\newblock \showarticletitle{What {D}oes {BERT} {L}ook {A}t? {A}n {A}nalysis of
  {BERT's} {A}ttention}. In \bibinfo{booktitle}{\emph{{BlackBoxNLP} @ {ACL}}}.
\newblock
\urldef\tempurl%
\url{http://arxiv.org/abs/1906.04341}
\showURL{%
\tempurl}


\bibitem[\protect\citeauthoryear{Cormack, Smucker, and Clarke}{Cormack
  et~al\mbox{.}}{2010}]%
        {Cormack2010EfficientAE}
\bibfield{author}{\bibinfo{person}{Gordon~V. Cormack}, \bibinfo{person}{Mark~D.
  Smucker}, {and} \bibinfo{person}{Charles L.~A. Clarke}.}
  \bibinfo{year}{2010}\natexlab{}.
\newblock \showarticletitle{Efficient and effective spam filtering and
  re-ranking for large web datasets}.
\newblock \bibinfo{journal}{\emph{Information Retrieval}}  \bibinfo{volume}{14}
  (\bibinfo{year}{2010}), \bibinfo{pages}{441--465}.
\newblock


\bibitem[\protect\citeauthoryear{Dai and Callan}{Dai and Callan}{2019}]%
        {Dai2019DeeperTU}
\bibfield{author}{\bibinfo{person}{Zhuyun Dai} {and} \bibinfo{person}{Jamie
  Callan}.} \bibinfo{year}{2019}\natexlab{}.
\newblock \showarticletitle{Deeper Text Understanding for IR with Contextual
  Neural Language Modeling}. In \bibinfo{booktitle}{\emph{SIGIR}}.
\newblock


\bibitem[\protect\citeauthoryear{Dai, Xiong, Callan, and Liu}{Dai
  et~al\mbox{.}}{2018}]%
        {dai-18}
\bibfield{author}{\bibinfo{person}{Zhuyun Dai}, \bibinfo{person}{Chenyan
  Xiong}, \bibinfo{person}{Jamie Callan}, {and} \bibinfo{person}{Zhiyuan Liu}.}
  \bibinfo{year}{2018}\natexlab{}.
\newblock \showarticletitle{Convolutional {N}eural {N}etworks for
  {Soft-Matching} {N-Grams} in {A}d-hoc {S}earch}. In
  \bibinfo{booktitle}{\emph{{WSDM}}}. \bibinfo{publisher}{{ACM} Press},
  \bibinfo{address}{Marina Del Rey, {CA}, {USA}}, \bibinfo{pages}{126--134}.
\newblock
\urldef\tempurl%
\url{http://dl.acm.org/citation.cfm?doid=3159652.3159659}
\showURL{%
\tempurl}


\bibitem[\protect\citeauthoryear{Dato, Lucchese, Nardini, Orlando, Perego,
  Tonellotto, and Venturini}{Dato et~al\mbox{.}}{2016}]%
        {tois:quickscorer}
\bibfield{author}{\bibinfo{person}{Domenico Dato}, \bibinfo{person}{Claudio
  Lucchese}, \bibinfo{person}{Franco~Maria Nardini}, \bibinfo{person}{Salvatore
  Orlando}, \bibinfo{person}{Raffaele Perego}, \bibinfo{person}{Nicola
  Tonellotto}, {and} \bibinfo{person}{Rossano Venturini}.}
  \bibinfo{year}{2016}\natexlab{}.
\newblock \showarticletitle{Fast Ranking with Additive Ensembles of Oblivious
  and Non-Oblivious Regression Trees}.
\newblock \bibinfo{journal}{\emph{ACM Transactions on Information Systems}}
  \bibinfo{volume}{35}, \bibinfo{number}{2} (\bibinfo{year}{2016}),
  \bibinfo{pages}{15:1--15:31}.
\newblock
\showISSN{1046-8188}


\bibitem[\protect\citeauthoryear{Dehghani, Zamani, Severyn, Kamps, and
  Croft}{Dehghani et~al\mbox{.}}{2017}]%
        {Dehghani2017NeuralRM}
\bibfield{author}{\bibinfo{person}{Mostafa Dehghani}, \bibinfo{person}{Hamed
  Zamani}, \bibinfo{person}{Aliaksei Severyn}, \bibinfo{person}{Jaap Kamps},
  {and} \bibinfo{person}{W.~Bruce Croft}.} \bibinfo{year}{2017}\natexlab{}.
\newblock \showarticletitle{Neural Ranking Models with Weak Supervision}. In
  \bibinfo{booktitle}{\emph{SIGIR}}.
\newblock


\bibitem[\protect\citeauthoryear{Devlin, Chang, Lee, and Toutanova}{Devlin
  et~al\mbox{.}}{2019}]%
        {devlin-19}
\bibfield{author}{\bibinfo{person}{Jacob Devlin}, \bibinfo{person}{{Ming-Wei}
  Chang}, \bibinfo{person}{Kenton Lee}, {and} \bibinfo{person}{Kristina
  Toutanova}.} \bibinfo{year}{2019}\natexlab{}.
\newblock \showarticletitle{{BERT}: {P}re-training of {D}eep {B}idirectional
  {T}ransformers for {L}anguage {U}nderstanding}. In
  \bibinfo{booktitle}{\emph{{NAACL}}}.
\newblock


\bibitem[\protect\citeauthoryear{Dietz and Gamari}{Dietz and Gamari}{2017}]%
        {Dietz2017}
\bibfield{author}{\bibinfo{person}{Laura Dietz} {and} \bibinfo{person}{Ben
  Gamari}.} \bibinfo{year}{2017}\natexlab{}.
\newblock \showarticletitle{{TREC CAR}: A Data Set for Complex Answer
  Retrieval}.
\newblock  (\bibinfo{year}{2017}).
\newblock
\urldef\tempurl%
\url{http://trec-car.cs.unh.edu}
\showURL{%
\tempurl}
\newblock
\shownote{Version 2.0.}


\bibitem[\protect\citeauthoryear{Guo, Fan, Ai, and Croft}{Guo
  et~al\mbox{.}}{2016}]%
        {guo-16}
\bibfield{author}{\bibinfo{person}{Jiafeng Guo}, \bibinfo{person}{Yixing Fan},
  \bibinfo{person}{Qingyao Ai}, {and} \bibinfo{person}{W.~Bruce Croft}.}
  \bibinfo{year}{2016}\natexlab{}.
\newblock \showarticletitle{A {D}eep {R}elevance {M}atching {M}odel for
  {A}d-hoc {R}etrieval}. In \bibinfo{booktitle}{\emph{{CIKM}}}.
  \bibinfo{pages}{55--64}.
\newblock
\urldef\tempurl%
\url{http://arxiv.org/abs/1711.08611}
\showURL{%
\tempurl}


\bibitem[\protect\citeauthoryear{Han, Mao, and Dally}{Han
  et~al\mbox{.}}{2015a}]%
        {Han2015DeepCC}
\bibfield{author}{\bibinfo{person}{Song Han}, \bibinfo{person}{Huizi Mao},
  {and} \bibinfo{person}{William~J. Dally}.} \bibinfo{year}{2015}\natexlab{a}.
\newblock \showarticletitle{Deep Compression: Compressing Deep Neural Network
  with Pruning, Trained Quantization and Huffman Coding}. In
  \bibinfo{booktitle}{\emph{ICLR}}.
\newblock


\bibitem[\protect\citeauthoryear{Han, Mao, and Dally}{Han
  et~al\mbox{.}}{2016}]%
        {DBLP:journals/corr/HanMD15}
\bibfield{author}{\bibinfo{person}{Song Han}, \bibinfo{person}{Huizi Mao},
  {and} \bibinfo{person}{William~J. Dally}.} \bibinfo{year}{2016}\natexlab{}.
\newblock \showarticletitle{Deep Compression: Compressing Deep Neural Network
  with Pruning, Trained Quantization and Huffman Coding}. In
  \bibinfo{booktitle}{\emph{4th International Conference on Learning
  Representations, {ICLR} 2016, San Juan, Puerto Rico, May 2-4, 2016,
  Conference Track Proceedings}}.
\newblock
\urldef\tempurl%
\url{http://arxiv.org/abs/1510.00149}
\showURL{%
\tempurl}


\bibitem[\protect\citeauthoryear{Han, Pool, Tran, and Dally}{Han
  et~al\mbox{.}}{2015b}]%
        {han2015learning}
\bibfield{author}{\bibinfo{person}{Song Han}, \bibinfo{person}{Jeff Pool},
  \bibinfo{person}{John Tran}, {and} \bibinfo{person}{William Dally}.}
  \bibinfo{year}{2015}\natexlab{b}.
\newblock \showarticletitle{Learning both weights and connections for efficient
  neural network}. In \bibinfo{booktitle}{\emph{Advances in neural information
  processing systems}}. \bibinfo{pages}{1135--1143}.
\newblock


\bibitem[\protect\citeauthoryear{Hendrycks and Gimpel}{Hendrycks and
  Gimpel}{2016}]%
        {hendrycks2016gaussian}
\bibfield{author}{\bibinfo{person}{Dan Hendrycks} {and} \bibinfo{person}{Kevin
  Gimpel}.} \bibinfo{year}{2016}\natexlab{}.
\newblock \showarticletitle{Gaussian error linear units (gelus)}.
\newblock \bibinfo{journal}{\emph{arXiv preprint arXiv:1606.08415}}
  (\bibinfo{year}{2016}).
\newblock


\bibitem[\protect\citeauthoryear{Hinton, Vinyals, and Dean}{Hinton
  et~al\mbox{.}}{2015}]%
        {DBLP:journals/corr/HintonVD15}
\bibfield{author}{\bibinfo{person}{Geoffrey~E. Hinton}, \bibinfo{person}{Oriol
  Vinyals}, {and} \bibinfo{person}{Jeffrey Dean}.}
  \bibinfo{year}{2015}\natexlab{}.
\newblock \showarticletitle{Distilling the Knowledge in a Neural Network}.
\newblock \bibinfo{journal}{\emph{CoRR}}  \bibinfo{volume}{abs/1503.02531}
  (\bibinfo{year}{2015}).
\newblock
\showeprint[arxiv]{1503.02531}
\urldef\tempurl%
\url{http://arxiv.org/abs/1503.02531}
\showURL{%
\tempurl}


\bibitem[\protect\citeauthoryear{Hofst{\"a}tter and Hanbury}{Hofst{\"a}tter and
  Hanbury}{2019}]%
        {Hofsttter2019LetsMR}
\bibfield{author}{\bibinfo{person}{Sebastian Hofst{\"a}tter} {and}
  \bibinfo{person}{Allan Hanbury}.} \bibinfo{year}{2019}\natexlab{}.
\newblock \showarticletitle{Let's measure run time! Extending the IR
  replicability infrastructure to include performance aspects}. In
  \bibinfo{booktitle}{\emph{OSIRRC@SIGIR}}.
\newblock


\bibitem[\protect\citeauthoryear{Huang, He, Gao, Deng, Acero, and Heck}{Huang
  et~al\mbox{.}}{2013}]%
        {Huang2013LearningDS}
\bibfield{author}{\bibinfo{person}{Po-Sen Huang}, \bibinfo{person}{Xiaodong
  He}, \bibinfo{person}{Jianfeng Gao}, \bibinfo{person}{Li Deng},
  \bibinfo{person}{Alex Acero}, {and} \bibinfo{person}{Larry~P. Heck}.}
  \bibinfo{year}{2013}\natexlab{}.
\newblock \showarticletitle{Learning deep structured semantic models for web
  search using clickthrough data}. In \bibinfo{booktitle}{\emph{CIKM}}.
\newblock


\bibitem[\protect\citeauthoryear{Hubara, Courbariaux, Soudry, El-Yaniv, and
  Bengio}{Hubara et~al\mbox{.}}{2017}]%
        {quantized2016}
\bibfield{author}{\bibinfo{person}{Itay Hubara}, \bibinfo{person}{Matthieu
  Courbariaux}, \bibinfo{person}{Daniel Soudry}, \bibinfo{person}{Ran
  El-Yaniv}, {and} \bibinfo{person}{Yoshua Bengio}.}
  \bibinfo{year}{2017}\natexlab{}.
\newblock \showarticletitle{Quantized neural networks: Training neural networks
  with low precision weights and activations}.
\newblock \bibinfo{journal}{\emph{The Journal of Machine Learning Research}}
  \bibinfo{volume}{18}, \bibinfo{number}{1} (\bibinfo{year}{2017}),
  \bibinfo{pages}{6869--6898}.
\newblock


\bibitem[\protect\citeauthoryear{Hui, Yates, Berberich, and de~Melo}{Hui
  et~al\mbox{.}}{2017}]%
        {Hui2017PACRRAP}
\bibfield{author}{\bibinfo{person}{Kai Hui}, \bibinfo{person}{Andrew Yates},
  \bibinfo{person}{Klaus Berberich}, {and} \bibinfo{person}{Gerard de Melo}.}
  \bibinfo{year}{2017}\natexlab{}.
\newblock \showarticletitle{PACRR: A Position-Aware Neural IR Model for
  Relevance Matching}. In \bibinfo{booktitle}{\emph{EMNLP}}.
\newblock


\bibitem[\protect\citeauthoryear{Huston and Croft}{Huston and Croft}{2014}]%
        {huston2014parameters}
\bibfield{author}{\bibinfo{person}{Samuel Huston} {and}
  \bibinfo{person}{W~Bruce Croft}.} \bibinfo{year}{2014}\natexlab{}.
\newblock \showarticletitle{Parameters learned in the comparison of retrieval
  models using term dependencies}.
\newblock \bibinfo{journal}{\emph{Technical Report}} (\bibinfo{year}{2014}).
\newblock


\bibitem[\protect\citeauthoryear{Ji, Shao, and Yang}{Ji et~al\mbox{.}}{2019}]%
        {ji-19}
\bibfield{author}{\bibinfo{person}{Shiyu Ji}, \bibinfo{person}{Jinjin Shao},
  {and} \bibinfo{person}{Tao Yang}.} \bibinfo{year}{2019}\natexlab{}.
\newblock \showarticletitle{Efficient {I}nteraction-based {N}eural {R}anking
  with {L}ocality {S}ensitive {H}ashing}. In \bibinfo{booktitle}{\emph{{WWW}}}.
\newblock


\bibitem[\protect\citeauthoryear{Jiao, Yin, Shang, Jiang, Chen, Li, Wang, and
  Liu}{Jiao et~al\mbox{.}}{2019}]%
        {Jiao2019TinyBERTDB}
\bibfield{author}{\bibinfo{person}{Xiaoqi Jiao}, \bibinfo{person}{Y. Yin},
  \bibinfo{person}{Lifeng Shang}, \bibinfo{person}{Xin Jiang},
  \bibinfo{person}{Xusong Chen}, \bibinfo{person}{Linlin Li},
  \bibinfo{person}{Fang Wang}, {and} \bibinfo{person}{Qun Liu}.}
  \bibinfo{year}{2019}\natexlab{}.
\newblock \showarticletitle{TinyBERT: Distilling BERT for Natural Language
  Understanding}.
\newblock \bibinfo{journal}{\emph{ArXiv}}  \bibinfo{volume}{abs/1909.10351}
  (\bibinfo{year}{2019}).
\newblock


\bibitem[\protect\citeauthoryear{Kingma and Ba}{Kingma and Ba}{2015}]%
        {Kingma2015AdamAM}
\bibfield{author}{\bibinfo{person}{Diederik~P. Kingma} {and}
  \bibinfo{person}{Jimmy Ba}.} \bibinfo{year}{2015}\natexlab{}.
\newblock \showarticletitle{Adam: {A} Method for Stochastic Optimization}. In
  \bibinfo{booktitle}{\emph{ICLR}}.
\newblock


\bibitem[\protect\citeauthoryear{Lettich, Lucchese, Nardini, Orlando, Perego,
  Tonellotto, and Venturini}{Lettich et~al\mbox{.}}{2018}]%
        {tpds}
\bibfield{author}{\bibinfo{person}{Francesco Lettich}, \bibinfo{person}{Claudio
  Lucchese}, \bibinfo{person}{Franco~Maria Nardini}, \bibinfo{person}{Salvatore
  Orlando}, \bibinfo{person}{Raffaele Perego}, \bibinfo{person}{Nicola
  Tonellotto}, {and} \bibinfo{person}{Rossano Venturini}.}
  \bibinfo{year}{2018}\natexlab{}.
\newblock \showarticletitle{Parallel Traversal of Large Ensembles of Decision
  Trees}.
\newblock \bibinfo{journal}{\emph{IEEE Transactions on Parallel and Distributed
  Systems}} (\bibinfo{year}{2018}), \bibinfo{pages}{14}.
\newblock
\showISSN{1045-9219}
\urldef\tempurl%
\url{https://doi.org/10.1109/TPDS.2018.2860982}
\showDOI{\tempurl}


\bibitem[\protect\citeauthoryear{Liu, Ott, Goyal, Du, Joshi, Chen, Levy, Lewis,
  Zettlemoyer, and Stoyanov}{Liu et~al\mbox{.}}{2019}]%
        {Liu2019RoBERTaAR}
\bibfield{author}{\bibinfo{person}{Yinhan Liu}, \bibinfo{person}{Myle Ott},
  \bibinfo{person}{Naman Goyal}, \bibinfo{person}{Jingfei Du},
  \bibinfo{person}{Mandar~S. Joshi}, \bibinfo{person}{Danqi Chen},
  \bibinfo{person}{Omer Levy}, \bibinfo{person}{Mike Lewis},
  \bibinfo{person}{Luke~S. Zettlemoyer}, {and} \bibinfo{person}{Veselin
  Stoyanov}.} \bibinfo{year}{2019}\natexlab{}.
\newblock \showarticletitle{RoBERTa: A Robustly Optimized BERT Pretraining
  Approach}.
\newblock \bibinfo{journal}{\emph{ArXiv}}  \bibinfo{volume}{abs/1907.11692}
  (\bibinfo{year}{2019}).
\newblock


\bibitem[\protect\citeauthoryear{MacAvaney}{MacAvaney}{2020}]%
        {macavaney:wsdm2020-onir}
\bibfield{author}{\bibinfo{person}{Sean MacAvaney}.}
  \bibinfo{year}{2020}\natexlab{}.
\newblock \showarticletitle{OpenNIR: A Complete Neural Ad-Hoc Ranking
  Pipeline}. In \bibinfo{booktitle}{\emph{WSDM}}.
\newblock


\bibitem[\protect\citeauthoryear{MacAvaney, Yates, Cohan, and
  Goharian}{MacAvaney et~al\mbox{.}}{2019a}]%
        {MacAvaney2019CEDRCE}
\bibfield{author}{\bibinfo{person}{Sean MacAvaney}, \bibinfo{person}{Andrew
  Yates}, \bibinfo{person}{Arman Cohan}, {and} \bibinfo{person}{Nazli
  Goharian}.} \bibinfo{year}{2019}\natexlab{a}.
\newblock \showarticletitle{CEDR: Contextualized Embeddings for Document
  Ranking}. In \bibinfo{booktitle}{\emph{SIGIR}}.
\newblock


\bibitem[\protect\citeauthoryear{MacAvaney, Yates, Hui, and Frieder}{MacAvaney
  et~al\mbox{.}}{2019b}]%
        {MacAvaney2019ContentBasedWS}
\bibfield{author}{\bibinfo{person}{Sean MacAvaney}, \bibinfo{person}{Andrew
  Yates}, \bibinfo{person}{Kai Hui}, {and} \bibinfo{person}{Ophir Frieder}.}
  \bibinfo{year}{2019}\natexlab{b}.
\newblock \showarticletitle{Content-Based Weak Supervision for Ad-Hoc
  Re-Ranking}. In \bibinfo{booktitle}{\emph{SIGIR}}.
\newblock


\bibitem[\protect\citeauthoryear{Matveeva, Burges, Burkard, Laucius, and
  Wong}{Matveeva et~al\mbox{.}}{2006}]%
        {Matveeva2006HighAR}
\bibfield{author}{\bibinfo{person}{Irina Matveeva},
  \bibinfo{person}{Christopher J.~C. Burges}, \bibinfo{person}{Timo Burkard},
  \bibinfo{person}{Andy Laucius}, {and} \bibinfo{person}{Leon Wong}.}
  \bibinfo{year}{2006}\natexlab{}.
\newblock \showarticletitle{High accuracy retrieval with multiple nested
  ranker}. In \bibinfo{booktitle}{\emph{SIGIR}}.
\newblock


\bibitem[\protect\citeauthoryear{Nanni, Mitra, Magnusson, and Dietz}{Nanni
  et~al\mbox{.}}{2017}]%
        {Nanni2017BenchmarkFC}
\bibfield{author}{\bibinfo{person}{Federico Nanni}, \bibinfo{person}{Bhaskar
  Mitra}, \bibinfo{person}{Matt Magnusson}, {and} \bibinfo{person}{Laura
  Dietz}.} \bibinfo{year}{2017}\natexlab{}.
\newblock \showarticletitle{Benchmark for Complex Answer Retrieval}. In
  \bibinfo{booktitle}{\emph{ICTIR}}.
\newblock


\bibitem[\protect\citeauthoryear{Nogueira and Cho}{Nogueira and Cho}{2019}]%
        {Nogueira2019PassageRW}
\bibfield{author}{\bibinfo{person}{Rodrigo Nogueira} {and}
  \bibinfo{person}{Kyunghyun Cho}.} \bibinfo{year}{2019}\natexlab{}.
\newblock \showarticletitle{Passage Re-ranking with BERT}.
\newblock \bibinfo{journal}{\emph{ArXiv}}  \bibinfo{volume}{abs/1901.04085}
  (\bibinfo{year}{2019}).
\newblock


\bibitem[\protect\citeauthoryear{Nogueira, Yang, Lin, and Cho}{Nogueira
  et~al\mbox{.}}{2019}]%
        {Nogueira2019DocumentEB}
\bibfield{author}{\bibinfo{person}{Rodrigo Nogueira}, \bibinfo{person}{Wei
  Yang}, \bibinfo{person}{Jimmy Lin}, {and} \bibinfo{person}{Kyunghyun Cho}.}
  \bibinfo{year}{2019}\natexlab{}.
\newblock \showarticletitle{Document Expansion by Query Prediction}.
\newblock \bibinfo{journal}{\emph{ArXiv}}  \bibinfo{volume}{abs/1904.08375}
  (\bibinfo{year}{2019}).
\newblock


\bibitem[\protect\citeauthoryear{Pan, Dong, and Guo}{Pan et~al\mbox{.}}{2016}]%
        {DBLP:journals/corr/PanDG16}
\bibfield{author}{\bibinfo{person}{Wei Pan}, \bibinfo{person}{Hao Dong}, {and}
  \bibinfo{person}{Yike Guo}.} \bibinfo{year}{2016}\natexlab{}.
\newblock \showarticletitle{DropNeuron: Simplifying the Structure of Deep
  Neural Networks}.
\newblock \bibinfo{journal}{\emph{CoRR}}  \bibinfo{volume}{abs/1606.07326}
  (\bibinfo{year}{2016}).
\newblock
\showeprint[arxiv]{1606.07326}
\urldef\tempurl%
\url{http://arxiv.org/abs/1606.07326}
\showURL{%
\tempurl}


\bibitem[\protect\citeauthoryear{Radford, Narasimhan, Salimans, and
  Sutskever}{Radford et~al\mbox{.}}{2018}]%
        {radford2018improving}
\bibfield{author}{\bibinfo{person}{Alec Radford}, \bibinfo{person}{Karthik
  Narasimhan}, \bibinfo{person}{Tim Salimans}, {and} \bibinfo{person}{Ilya
  Sutskever}.} \bibinfo{year}{2018}\natexlab{}.
\newblock \bibinfo{booktitle}{\emph{Improving language understanding by
  generative pre-training}}.
\newblock \bibinfo{type}{{T}echnical {R}eport}. \bibinfo{institution}{OpenAI}.
\newblock


\bibitem[\protect\citeauthoryear{Rosset, Jose, Ghosh, Mitra, and Tiwary}{Rosset
  et~al\mbox{.}}{2018}]%
        {Rosset2018OptimizingQE}
\bibfield{author}{\bibinfo{person}{Corby Rosset}, \bibinfo{person}{Damien
  Jose}, \bibinfo{person}{Gargi Ghosh}, \bibinfo{person}{Bhaskar Mitra}, {and}
  \bibinfo{person}{Saurabh Tiwary}.} \bibinfo{year}{2018}\natexlab{}.
\newblock \showarticletitle{Optimizing Query Evaluations Using Reinforcement
  Learning for Web Search}. In \bibinfo{booktitle}{\emph{SIGIR}}.
\newblock


\bibitem[\protect\citeauthoryear{Sanh, Debut, Chaumond, and Wolf}{Sanh
  et~al\mbox{.}}{2019}]%
        {Sanh2019DistilBERTAD}
\bibfield{author}{\bibinfo{person}{Victor Sanh}, \bibinfo{person}{Lysandre
  Debut}, \bibinfo{person}{Julien Chaumond}, {and} \bibinfo{person}{Thomas
  Wolf}.} \bibinfo{year}{2019}\natexlab{}.
\newblock \showarticletitle{DistilBERT, a distilled version of BERT: smaller,
  faster, cheaper and lighter}. In \bibinfo{booktitle}{\emph{Workshop on Energy
  Efficient Machine Learning and Cognitive Computing @ NeuIPS}}.
\newblock


\bibitem[\protect\citeauthoryear{Schwartz, Dodge, Smith, and Etzioni}{Schwartz
  et~al\mbox{.}}{2019}]%
        {Schwartz2019GreenA}
\bibfield{author}{\bibinfo{person}{Roy Schwartz}, \bibinfo{person}{Jesse
  Dodge}, \bibinfo{person}{Noah~A. Smith}, {and} \bibinfo{person}{Oren
  Etzioni}.} \bibinfo{year}{2019}\natexlab{}.
\newblock \showarticletitle{Green AI}.
\newblock \bibinfo{journal}{\emph{ArXiv}}  \bibinfo{volume}{abs/1907.10597}
  (\bibinfo{year}{2019}).
\newblock


\bibitem[\protect\citeauthoryear{Seo and Kim}{Seo and Kim}{2019}]%
        {Seo2019EfficientWQ}
\bibfield{author}{\bibinfo{person}{Sanghyun Seo} {and} \bibinfo{person}{Juntae
  Kim}.} \bibinfo{year}{2019}\natexlab{}.
\newblock \showarticletitle{Efficient Weights Quantization of Convolutional
  Neural Networks Using Kernel Density Estimation based Non-uniform Quantizer}.
\newblock \bibinfo{journal}{\emph{Appl. Sci}} (\bibinfo{year}{2019}).
\newblock


\bibitem[\protect\citeauthoryear{Shoeybi, Patwary, Puri, LeGresley, Casper, and
  Catanzaro}{Shoeybi et~al\mbox{.}}{2019}]%
        {Shoeybi2019MegatronLMTM}
\bibfield{author}{\bibinfo{person}{Mohammad Shoeybi},
  \bibinfo{person}{Mostofa~Ali Patwary}, \bibinfo{person}{Raul Puri},
  \bibinfo{person}{Patrick LeGresley}, \bibinfo{person}{Jared Casper}, {and}
  \bibinfo{person}{Bryan Catanzaro}.} \bibinfo{year}{2019}\natexlab{}.
\newblock \showarticletitle{Megatron-LM: Training Multi-Billion Parameter
  Language Models Using Model Parallelism}.
\newblock \bibinfo{journal}{\emph{ArXiv}}  \bibinfo{volume}{abs/1909.08053}
  (\bibinfo{year}{2019}).
\newblock


\bibitem[\protect\citeauthoryear{Tang, Lu, Liu, Mou, Vechtomova, and Lin}{Tang
  et~al\mbox{.}}{2019}]%
        {Tang2019DistillingTK}
\bibfield{author}{\bibinfo{person}{Raphael Tang}, \bibinfo{person}{Yao Lu},
  \bibinfo{person}{Linqing Liu}, \bibinfo{person}{Lili Mou},
  \bibinfo{person}{Olga Vechtomova}, {and} \bibinfo{person}{Jimmy Lin}.}
  \bibinfo{year}{2019}\natexlab{}.
\newblock \showarticletitle{Distilling Task-Specific Knowledge from BERT into
  Simple Neural Networks}.
\newblock \bibinfo{journal}{\emph{ArXiv}}  \bibinfo{volume}{abs/1903.12136}
  (\bibinfo{year}{2019}).
\newblock


\bibitem[\protect\citeauthoryear{Tonellotto, Macdonald, and Ounis}{Tonellotto
  et~al\mbox{.}}{2018}]%
        {fnt}
\bibfield{author}{\bibinfo{person}{Nicola Tonellotto}, \bibinfo{person}{Craig
  Macdonald}, {and} \bibinfo{person}{Iadh Ounis}.}
  \bibinfo{year}{2018}\natexlab{}.
\newblock \showarticletitle{Efficient Query Processing for Scalable Web
  Search}.
\newblock \bibinfo{journal}{\emph{Foundations and Trends in Information
  Retrieval}} \bibinfo{volume}{12}, \bibinfo{number}{4--5}
  (\bibinfo{year}{2018}), \bibinfo{pages}{319--492}.
\newblock
\showISSN{1554-0669}


\bibitem[\protect\citeauthoryear{Vaswani, Shazeer, Parmar, Uszkoreit, Jones,
  Gomez, Kaiser, and Polosukhin}{Vaswani et~al\mbox{.}}{2017}]%
        {vaswani-17}
\bibfield{author}{\bibinfo{person}{Ashish Vaswani}, \bibinfo{person}{Noam
  Shazeer}, \bibinfo{person}{Niki Parmar}, \bibinfo{person}{Jakob Uszkoreit},
  \bibinfo{person}{Llion Jones}, \bibinfo{person}{Aidan~N. Gomez},
  \bibinfo{person}{Lukasz Kaiser}, {and} \bibinfo{person}{Illia Polosukhin}.}
  \bibinfo{year}{2017}\natexlab{}.
\newblock \showarticletitle{Attention {I}s {A}ll {Y}ou {N}eed}. In
  \bibinfo{booktitle}{\emph{{NeuIPS}}}.
\newblock
\urldef\tempurl%
\url{http://arxiv.org/abs/1706.03762}
\showURL{%
\tempurl}


\bibitem[\protect\citeauthoryear{Wang, Lin, and Metzler}{Wang
  et~al\mbox{.}}{2011}]%
        {wang2011cascade}
\bibfield{author}{\bibinfo{person}{Lidan Wang}, \bibinfo{person}{Jimmy Lin},
  {and} \bibinfo{person}{Donald Metzler}.} \bibinfo{year}{2011}\natexlab{}.
\newblock \showarticletitle{A cascade ranking model for efficient ranked
  retrieval}. In \bibinfo{booktitle}{\emph{SIGIR}}.
\newblock


\bibitem[\protect\citeauthoryear{Xiong, Dai, Callan, Liu, and Power}{Xiong
  et~al\mbox{.}}{2017}]%
        {xiong-17}
\bibfield{author}{\bibinfo{person}{Chenyan Xiong}, \bibinfo{person}{Zhuyun
  Dai}, \bibinfo{person}{Jamie Callan}, \bibinfo{person}{Zhiyuan Liu}, {and}
  \bibinfo{person}{Russell Power}.} \bibinfo{year}{2017}\natexlab{}.
\newblock \showarticletitle{{End-to-End} {N}eural {A}d-hoc {R}anking with
  {K}ernel {P}ooling}. In \bibinfo{booktitle}{\emph{{SIGIR}}}.
  \bibinfo{pages}{55--64}.
\newblock
\urldef\tempurl%
\url{http://arxiv.org/abs/1706.06613}
\showURL{%
\tempurl}
\newblock
\shownote{{arXiv}: 1706.06613.}


\bibitem[\protect\citeauthoryear{{Xu}, {Yao}, {Lin}, {Ou}, {Cao}, {Wang}, and
  {Zha}}{{Xu} et~al\mbox{.}}{2018}]%
        {2018alternating}
\bibfield{author}{\bibinfo{person}{Chen {Xu}}, \bibinfo{person}{Jianqiang
  {Yao}}, \bibinfo{person}{Zhouchen {Lin}}, \bibinfo{person}{Wenwu {Ou}},
  \bibinfo{person}{Yuanbin {Cao}}, \bibinfo{person}{Zhirong {Wang}}, {and}
  \bibinfo{person}{Hongbin {Zha}}.} \bibinfo{year}{2018}\natexlab{}.
\newblock \showarticletitle{Alternating Multi-bit Quantization for Recurrent
  Neural Networks}. In \bibinfo{booktitle}{\emph{International Conference on
  Learning Representations}}.
\newblock
\urldef\tempurl%
\url{https://arxiv.org/abs/1802.00150}
\showURL{%
\tempurl}


\bibitem[\protect\citeauthoryear{Yang, Fang, and Lin}{Yang
  et~al\mbox{.}}{2017}]%
        {Yang2017AnseriniET}
\bibfield{author}{\bibinfo{person}{Peilin Yang}, \bibinfo{person}{Hui Fang},
  {and} \bibinfo{person}{Jimmy Lin}.} \bibinfo{year}{2017}\natexlab{}.
\newblock \showarticletitle{Anserini: Enabling the Use of Lucene for
  Information Retrieval Research}. In \bibinfo{booktitle}{\emph{SIGIR}}.
\newblock


\bibitem[\protect\citeauthoryear{Yang, Xie, Lin, Li, Tan, Xiong, Li, and
  Lin}{Yang et~al\mbox{.}}{2019a}]%
        {Yang2019EndtoEndOQ}
\bibfield{author}{\bibinfo{person}{Wei Yang}, \bibinfo{person}{Yuqing Xie},
  \bibinfo{person}{Aileen Lin}, \bibinfo{person}{Xingyu Li},
  \bibinfo{person}{Luchen Tan}, \bibinfo{person}{Kun Xiong},
  \bibinfo{person}{Ming Li}, {and} \bibinfo{person}{Jimmy Lin}.}
  \bibinfo{year}{2019}\natexlab{a}.
\newblock \showarticletitle{End-to-End Open-Domain Question Answering with
  BERTserini}. In \bibinfo{booktitle}{\emph{NAACL-HLT}}.
\newblock


\bibitem[\protect\citeauthoryear{Yang, Zhang, and Lin}{Yang
  et~al\mbox{.}}{2019b}]%
        {Yang2019SimpleAO}
\bibfield{author}{\bibinfo{person}{Wei Yang}, \bibinfo{person}{Haotian Zhang},
  {and} \bibinfo{person}{Jimmy Lin}.} \bibinfo{year}{2019}\natexlab{b}.
\newblock \showarticletitle{Simple Applications of BERT for Ad Hoc Document
  Retrieval}.
\newblock \bibinfo{journal}{\emph{ArXiv}}  \bibinfo{volume}{abs/1903.10972}
  (\bibinfo{year}{2019}).
\newblock


\bibitem[\protect\citeauthoryear{Yilmaz, Wang, Yang, Zhang, and Lin}{Yilmaz
  et~al\mbox{.}}{2019}]%
        {Yilmaz2019ApplyingBT}
\bibfield{author}{\bibinfo{person}{Zeynep~Akkalyoncu Yilmaz},
  \bibinfo{person}{Shengjin Wang}, \bibinfo{person}{Wei Yang},
  \bibinfo{person}{Haotian Zhang}, {and} \bibinfo{person}{Jimmy Lin}.}
  \bibinfo{year}{2019}\natexlab{}.
\newblock \showarticletitle{Applying BERT to Document Retrieval with Birch}. In
  \bibinfo{booktitle}{\emph{EMNLP/IJCNLP}}.
\newblock


\bibitem[\protect\citeauthoryear{Zamani, Dehghani, Croft, Learned-Miller, and
  Kamps}{Zamani et~al\mbox{.}}{2018}]%
        {zamani2018neural}
\bibfield{author}{\bibinfo{person}{Hamed Zamani}, \bibinfo{person}{Mostafa
  Dehghani}, \bibinfo{person}{W~Bruce Croft}, \bibinfo{person}{Erik
  Learned-Miller}, {and} \bibinfo{person}{Jaap Kamps}.}
  \bibinfo{year}{2018}\natexlab{}.
\newblock \showarticletitle{From neural re-ranking to neural ranking: Learning
  a sparse representation for inverted indexing}. In
  \bibinfo{booktitle}{\emph{CIKM}}.
\newblock


\end{thebibliography}

\end{document}